\documentclass[preprint,pteplogo]{ptephy}%%%%%% to generate preprint number with ptep logo

\preprintnumber{KUNS2618} %%% Insert preprint number here

\newcommand{\Slash}[1]{{\ooalign{\hfil/\hfil\crcr$#1$}}}
\newcommand{\tr}{{\rm tr}}

\newcommand{\im}{{\rm Im}}
\newcommand{\sigu}{{\left<\sigma_u\right>}}
\newcommand{\sigd}{{\left<\sigma_d\right>}}
\newcommand{\sigs}{{\left<\sigma_s\right>}}

\newcommand{\sigq}{{\left<\sigma_q\right>}}
\newcommand{\dsq}{{\left<\delta\sigma_q\right>}}
\newcommand{\ppp}{{\pi^+\pi^-\pi^0}}
\newcommand{\tpe}{{\theta_{\pi^0\eta}}}
\newcommand{\tpep}{{\theta_{\pi^0\eta'}}}

\newcommand{\tss}{{\theta_{\sigma_3\sigma}}}
\newcommand{\tsf}{{\theta_{\sigma_3f_0}}}
\newcommand{\tps}{{\theta_{ps}}}
\newcommand{\ts}{{\theta_s}}

\newcommand{\eppp}{{\eta\rightarrow\pi^+\pi^-\pi^0}}

\newcommand{\qt}{{\tilde{q}}}
\newcommand{\dsigq}{{\delta\left<\sigma_q\right>}}
\newcommand{\dsigs}{{\delta\left<\sigma_s\right>}}

\usepackage{amsmath}	% required for `\align' (yatex added)
\usepackage{graphicx}	% required for `\includegraphics' (yatex added)
\usepackage{ulem}
\allowdisplaybreaks
\begin{document}

\title{Possible relevance of the softening of the sigma meson to
$\eta$ decay into $3\pi$ in the nuclear medium}

\author{\name{Shuntaro Sakai}{} and \name{Teiji Kunihiro}{}}
%%%%%%%%%%% The \name command should be used as \name{Insert author name here}{Insert affiliation number here}
%%%%% Please use \thanks for contributed author details

%%%%%%%%%%% The \affil command should be used as \affil{Insert affiliation number here}{Insert author address here}
\address{\affil{}{Department of Physics, Kyoto University,
Kitashirakawa-Oiwakecho, Kyoto 606-8502, Japan}
%\affil{2}{Second author address}
%\affil{3}{Third author address}
\email{s.sakai@ruby.kyoto-u.ac.jp}}

\begin{abstract}%
 We investigate the role
 %for
 of the softening of the scalar$-$isoscalar
 (sigma) meson in the $\eta\rightarrow\ppp$ and $3\pi^0$ decay widths in
 the symmetric nuclear medium using a linear sigma model.
% The softening of the sigma meson occurs associated with the chiral
% restoration within the model.
 Our calculation shows that these decay widths in the nuclear medium
 increase
 by up to a factor of four to ten compared with those in the free space
 mainly depending on the mass of the sigma meson in the free space which is an
 input parameter of the model.
% The softening of the sigma meson gives a significant effect on the
% enhancements.
 The enhancements are considerable even at a half of the normal nuclear
 density.
 Thus, the $\eta$ decay into $3\pi$
 %can be a new possible probe
could be a possible new probe for the chiral restoration in the nuclear
 medium.
% The ratio of the decay widths of $\eta\rightarrow\ppp$ to $3\pi^0$
% decreases around the density where the in-medium $\eta\rightarrow3\pi$
% decay widths are most enhanced.
% We find that the nuclear-medium effect on the $\eta\rightarrow3\pi^0$
% decay width is more moderate than that on the $\eta\rightarrow\ppp$
 We find that the density dependence of the $\eta\rightarrow3\pi^0$
 decay is moderate in comparison with that of
%the
 $\eppp$,
% This is because of the cancellation of the terms appearing from
% the crossing symmetry.
 although the former width is
 greater than the latter one at a given density:
 This is because the softening of the sigma meson causes
% the
 cancellation of the terms appearing from the Bose symmetry in the
 $\eta\rightarrow3\pi^0$ decay.
 The difference
 % of
 between the density dependences should be 
 helpful for
 %an
 experimental confirmation of the findings of the present study.
 \end{abstract}

\subjectindex{xxxx, xxx}

\maketitle

\section{Introduction\label{intro}}
The $\eta\rightarrow3\pi$ decay process\footnote{We denote
the decay processes $\eta\rightarrow\ppp$ and $3\pi^0$
%as
by
$\eta\rightarrow3\pi$ for short.}
is prohibited by
%the
$G$ parity conservation,
which,
conversely,
implies that the isospin-symmetry breaking (ISB)
%, conversely,
can cause the process.
The possible sources of ISB are the electromagnetic
interaction and the tiny mass difference of the $u$ and $d$ quarks from the viewpoint of quantum chromodynamics.
%The studies
Studies using the current-algebra technique
\cite{sutherland1966current,PhysRev.161.1483,bell1968current,osborn1970eta},
however, revealed that
%the
their
leading-order
contributions
%of them only
give a far smaller value of the decay width
than the experimental one.
The smallness of the higher-order contributions from the electromagnetic
effect is confirmed in
Refs.~\cite{baur1996electromagnetic,ditsche2009electromagnetic}.
There
%are
have been
many investigations based on the strong interaction
\cite{schechter1971delta,hudnall1974eta,weinberg1975u,kogut1975quark,raby1976calculation,kawarabayashi1980eta,leutwyler1996implications}\footnote{
%The works
Works involving the scalar meson, which had been previously reported in,
%for example,
e.g., Ref.~\cite{RevModPhys.36.977},
%is
are
summarized in
Ref.~\cite{salvini1969eta}.} because it should be
responsible for the discrepancy of the decay widths between the
experimental value and the current-algebra estimation.
In addition to ISB,
the significant effect of the final-state interaction (FSI) was pointed
out in Refs.~\cite{neveu1970final,roiesnel1981resolution};
see
Refs.~\cite{gasser1985eta,kambor1996final,Anisovich1996,abdel2003effects,borasoy2005hadronic,bijnens2007eta,lanz2013eta,guo:2015zqa}
for later studies on this subject.

The modification of
%the
hadron properties in the nuclear medium
%has been one of the interesting topics in the
is an interesting topic in
hadron physics.
A focus of the study is to investigate
%the
chiral restoration in the
nuclear medium (see,
% for example
e.g., Ref.~\cite{hayano2010hadron}
%as
for a summary of the theoretical and experimental status);
the chiral restoration means the reduction of the quark condensate which
is the order parameter of the spontaneous breaking of chiral symmetry,
in the environment
%(for example,
(e.g., a heat bath or hadronic matter).

In the previous study \cite{sakai2014etadecay}, the present
authors investigated the
$\eta\rightarrow3\pi$ decay width in the asymmetric nuclear medium,
anticipating that  
the external isospin asymmetry would enhance the decay width;
the isospin symmetry is explicitly broken by the asymmetric nuclear
density $\delta\rho=\rho_n-\rho_p\neq0$ besides the mass difference of
the $u$ and $d$ quarks, where $\rho_p$ and $\rho_n$ are the proton and
neutron number densities, respectively.
The analysis is based on a nonlinear sigma model.
The effect of FSI in the $I=J=0$ channel, i.e.
the sigma-meson channel is included within the meson one-loop diagrams
\cite{gasser1985eta,bijnens2007eta}.
Then, the decay width in the free space shows a fairly good agreement with
the experimental value
and the remaining minor effects may be accounted for by a more complete
treatment of FSI,
such as the incorporation of the $\rho$ meson contribution in the t-channel.
It was found that the $\eta\rightarrow3\pi$ decay width is
enhanced as the total baryon density $\rho=\rho_n+\rho_p$
as well as $\delta\rho$ increases.
In the enhancement of the decay width by $\rho$,
the 4-meson--$N$--$N$ vertex, which is active only in the nuclear medium,
gives a significant contribution,
as in the enhancement of the in-medium $\pi\pi$
cross section in the $I=J=0$ channel \cite{jido2000medium}.
The $4\pi$--$N$--$N$ vertex is traced back to the contribution from the
scalar meson in the linear representation \cite{hatsuda1999precursor}.
Then, we expect
that the enhancement of the $\eta\rightarrow3\pi$ decay
width is also associated with the dynamics of the scalar meson in the
nuclear medium.

The nature of
%the
light scalar mesons has attracted
much attention in recent years; 
the observed resonance in the
$\pi\pi$ ($I=J=0$) channel 
(see,
%for example,
e.g., the section ``Note on scalar meson below 2 GeV'' in
Ref.~\cite{Agashe:2014kda} for a summary of the current status)
may be a hadronic
molecule of $2\pi$ \cite{oller1997chiral},
a tetraquark meson \cite{jaffe1977multiquark}, or a light scalar
glueball \cite{PhysRevD.33.801}.
In addition, it can be a quantum fluctuation of
the order parameter associated with the spontaneous breaking of the
chiral symmetry
\cite{nambu1961dynamical1,nambu1961dynamical2,hatsuda1994qcd}.
The observed scalar meson would be
%the
a superposition of the 
aforementioned states.
For the scalar meson in L$\sigma$M which is the fluctuation mode of
the order parameter and
is
called the sigma meson in the following, it is
anticipated to be softened through
%the
chiral restoration
\cite{Hatsuda:1985ey,PhysRevLett.55.158,Bernard:1987im,hatsuda1994qcd,Kunihiro:1995rb}.
 Recalling that FSI in the sigma-meson channel makes a significant
contribution,
we expect that the $\eta\rightarrow3\pi$ decay width is enhanced
in association with the softening of the sigma meson, and  thereby
the chiral restoration in the nuclear medium can be probed through the decay process.

\begin{figure}[t]
 \centering
 \includegraphics[width=6cm]{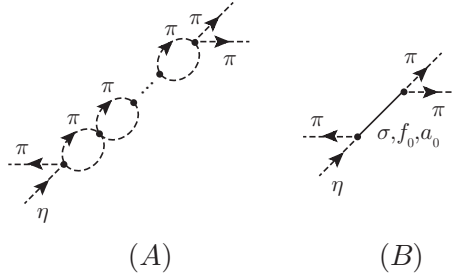}
 \caption{
 %The
 A schematic figure
 %of
 showing the
 FSI of pions.
 The solid and dashed lines
 %mean
 indicate
 the scalar and pseudoscalar
 mesons.
% The diagram $(A)$ represents the rescattering of pions in the final
% state.
 In this study, we assume that the contribution from the poles of the
 scalar mesons given in
 %the
 diagram $(B)$ can be substituted for the
 effect of the FSI of pions represented as
 %the
 diagram $(A)$.}
 \label{fig_FSI}
\end{figure}

In this study, we investigate the $\eta\rightarrow3\pi$ decay width in the
symmetric nuclear medium using L$\sigma$M, focusing on
the softening of the sigma meson.
Within L$\sigma$M, the softening of the sigma meson is automatically
built in.
As alluded to above, the formation of the resonance is responsible for FSI in
the $\pi\pi$ $(I=J=0)$ channel;
%: See
see Fig.~\ref{fig_FSI} for a schematic figure of FSI,
where
the $\rho$ meson contribution in the $\pi\pi$ $(I=1)$ channel is ignored
%expecting
in the expectation
that the contribution is small
because its pole position
is far from the
kinematically allowed region of the Dalitz plot of the
$\eta\rightarrow3\pi$ decay.
The in-medium modification of the $\rho$ meson properties,
%for example,
e.g., the mass reduction \cite{Brown:1995qt} or the broadening of the
width \cite{Rapp:1999ej}, would affect the $\eta\rightarrow3\pi$ decay in
the nuclear medium, though its contribution will be ignored, as stated above:
The modification of the $\rho$ meson properties in the nuclear medium
would also be small because our calculation is limited
%in
to
the
small
%density,
densities,
%as it
as
will be clarified in the following section.
%Then,
Thus,
the effect from the $\rho$ meson in the nuclear medium would be
small
in the situation treated in the present work.

It is known that the couplings with
%the
excited baryons, e.g.
$N^\ast(1535)$,
%give the
have interesting effects on the in-medium
properties of the $\eta$ or $\pi$ mesons (see,
%for example,
e.g.,
Ref.~\cite{kelkar2013interaction}).
%They
These
are not taken into account in the present work either,
%for focusing
to allow us to focus
on the role
%for
of
the
softening of the sigma meson in the $\eta\rightarrow3\pi$ decay.
The inclusion of these effects will be left
%as a future task.
for future work.

In the present paper, we shall show an enhancement of the
$\eta\rightarrow3\pi$ decay width in the nuclear medium.
The decay width becomes about four to ten times larger
than the value in the free space depending on the mass of the
sigma meson in the free space, which is an input
parameter
of our calculation.
The mechanism of the resultant enhancement is found to be similar to that
of the $\pi\pi(I=J=0)$
cross section in the nuclear medium \cite{hatsuda1999precursor}, where
the growth of the spectral function in the sigma channel near the $2\pi$
threshold plays an essential role.
The enhancement of the decay width is considerable even
%in the small density: Indeed,
at small densities; indeed,
the decay width at
%a
half of the normal nuclear density
is several times larger than that in the free space.

We find that the enhancement of the $\eta\rightarrow3\pi^0$
decay width in the nuclear medium is weaker than
that of the $\eppp$ decay at large $\rho$ though the decay
width of
$\eta\rightarrow3\pi^0$ itself is larger than that of $\eppp$. 
We shall show that this is because
%a part
some
of the terms appearing from
the Bose symmetry of $3\pi^0$
%are cancelled with
cancel
each other when the
mass of the sigma meson approaches the $2\pi$ threshold.

This paper is constructed as follows.
In Sect.~\ref{sec_setup}, we introduce the setup of our model.
In the subsequent section, we explain the method
%to take in
used to include
the effect of the
nuclear medium.
Then in Sect.~\ref{sec_result}, we present the numerical results of the
decay widths of the $\eta$ meson into $\ppp$ and $3\pi^0$ with some
discussions.
Finally, we give a summary and some remarks in Sect.~\ref{sec_conclusion}.
Some details of our calculation are presented in the appendices.

\section{Model\label{sec_setup}}
In this section, we introduce the model used in our calculation.

The meson and baryon fields belong to the $({\bf 3}_L,\bar{\bf
3}_R)\oplus(\bar{\bf 3}_L,{\bf 3}_R)$ representation of the chiral SU(3)
group.
The meson (baryon) field $M (B)$ is transformed into $LM(B)R^\dagger$
under the chiral SU(3) transformation, where $L,R$ are
%the
elements of
SU(3).
For the baryon field, we
omit the hyperons which do not appear 
in this calculation.

The Lagrangian of L$\sigma$M used in this study is given as follows:
%;
\begin{align}
 \mathcal{L}=&\frac{1}{2}\tr(\partial_\mu M\partial^\mu
 M^\dagger)-\frac{\mu^2}{2}\tr(MM^\dagger)-\frac{\lambda}{4}\tr(MM^\dagger)^2-\frac{\lambda'}{4}[\tr
 (MM^\dagger)]^2+\frac{B}{2}(\det M+\det M^\dagger)\notag\\
 &+\frac{A}{2}\tr(\chi
 M^\dagger+M\chi^\dagger)+\bar{N}\left(i\Slash{\partial}-gM_5\right)N,\label{eq_lsm_lagrangian} \\
 M=&M_{s}+iM_{ps}=\sum_{a=0}^8\frac{\lambda_a\sigma_a}{\sqrt{2}}+i\sum_{a=0}^8\frac{\lambda_a\pi_a}{\sqrt{2}},\
 N=^t(p,n),\ \chi={\rm diag}(m_u,m_d,m_s),\\
 M_{s}=&
 \begin{pmatrix}
  \sigma_u&a_0^+&\kappa^+\\
  a_0^-&\sigma_d&\kappa^0 \\
  \kappa^-&\bar{\kappa}^0 &\sigma_s 
 \end{pmatrix},\\
 M_{ps}=&
 \begin{pmatrix}
  \eta_0/\sqrt{3}+\pi_3/\sqrt{2}+\eta_8/\sqrt{6}&\pi^+&K^+\\
  \pi^-&\eta_0/\sqrt{3}-\pi_3/\sqrt{2}+\eta_8/\sqrt{6} &K^0 \\
  K^-&\bar{K}^0 &\eta_0/\sqrt{3}-2\eta_8/\sqrt{6} 
 \end{pmatrix},\\ 
 M_5=&
 \begin{pmatrix}
  \sigma_u&\\
  &\sigma_d 
 \end{pmatrix}
 +i\gamma_5
 \begin{pmatrix}
  \frac{\eta_0}{\sqrt{3}}+\frac{\pi_3}{\sqrt{2}}+\frac{\eta_8}{\sqrt{6}}&\\
  & \frac{\eta_0}{\sqrt{3}}-\frac{\pi_3}{\sqrt{2}}+\frac{\eta_8}{\sqrt{6}} 
 \end{pmatrix},\\
 \sigma_u=&\frac{\sigma_0}{\sqrt{3}}+\frac{\sigma_3}{\sqrt{2}}+\frac{\sigma_8}{\sqrt{6}},\
 \sigma_d=\frac{\sigma_0}{\sqrt{3}}-\frac{\sigma_3}{\sqrt{2}}+\frac{\sigma_8}{\sqrt{6}},\
 \sigma_s=\frac{\sigma_0}{\sqrt{3}}-\frac{2}{\sqrt{6}}\sigma_8. 
\end{align}
Here, $\lambda^a$ is the Gell-Mann matrix with the normalization
$\tr(\lambda^a\lambda^b)=2\delta^{ab}$.

The tree-level effective potential of the scalar fields in the nuclear medium
$V_{\rm eff}(\sigma_u,\sigma_d,\sigma_s;\rho)$ and the conditions to
determine their expectation values are given as follows;
\begin{align}
 V_{\rm eff}(\sigma_u,\sigma_d,\sigma_s;\rho)=&\frac{\mu^2}{2}(\sigma_u^2+\sigma_d^2+\sigma_s^2)+\frac{\lambda}{4}(\sigma_u^4+\sigma_d^4+\sigma_s^4)+\frac{\lambda'}{4}(\sigma_u^2+\sigma_d^2+\sigma_s^2)^2\notag\\
 &-B\sigma_u\sigma_d\sigma_s-A(m_u\sigma_u+m_d\sigma_d+m_s\sigma_s)+g(\rho_p+\rho_n), \\
 \frac{\partial V_{\rm eff}}{\partial
 \sigma_u}=&\mu^2\sigma_u+\lambda\sigma_u^3+\lambda'\sigma_u(\sigma_u^2+\sigma_d^2+\sigma_s^2)-B\sigma_d\sigma_s-Am_u+g\rho_p=0,\label{vc1}\\
 \frac{\partial V_{\rm eff}}{\partial
 \sigma_d}=&\mu^2\sigma_d+\lambda\sigma_d^3+\lambda'\sigma_d(\sigma_u^2+\sigma_d^2+\sigma_s^2)-B\sigma_u\sigma_s-Am_d+g\rho_n=0,\label{vc2}\\
 \frac{\partial V_{\rm eff}}{\partial
 \sigma_s}=&\mu^2\sigma_s+\lambda\sigma_s^3+\lambda'\sigma_s(\sigma_u^2+\sigma_d^2+\sigma_s^2)-B\sigma_u\sigma_d-Am_s=0.\label{vc3}
\end{align}
%\sout{ where $\rho_p$ and $\rho_n$ are the proton and the neutron number
%density.}
Using the Fermi momentum of
the
proton (neutron) $k_f^{p(n)}$,
the
proton (neutron)
number density
% \sout{$\rho_{p(n)}$} 
is expressed as
$\rho_{p(n)}=k_f^{p(n)3}/3\pi^2$.
Here, we assume
%the only
only the
scalar fields to have the nonzero expectation
values.
The expectation values of the sigma fields $\sigu,\sigd$, and $\sigs$ are
determined so as to solve these equations.
We keep only the terms up to $O(k_f^3)$, which corresponds to the Fermi
gas approximation.
Then, the contribution from the Fermi sea is omitted because it is
$O(k_f^5)$ within the nonrelativistic approximation for the nucleon
field.
The effect of the nucleon loop in the free space is also neglected in this
calculation
%expecting
in the expectation
that the effect is suppressed by the large nucleon
mass.

From the Lagrangian given by Eq.~(\ref{eq_lsm_lagrangian}), the axial
current $A_\mu^a$ is obtained as
\begin{align}
 A_\mu^a=&\tr\left[\partial_\mu
 M_{ps}\left\{\lambda^a,M_s\right\}-\partial_\mu
 M_s\left\{\lambda^a,M_{ps}\right\}\right].
\end{align}
Comparing
this
with the definition of the meson decay constants,
$\left<0\left|A_\mu^a(x)\right|\pi^b(p)\right>=ip_\mu
f_a\delta^{ab}e^{-ip\cdot x}$,
%the
those of the pion and the
kaon are obtained as
\begin{align}
 f_{\pi^0}=f_{\pi^\pm}=&\frac{\sigu+\sigd}{\sqrt{2}},\label{eq_fpi}\\
 f_{K^\pm}=&\frac{\sigu+\sigs}{\sqrt{2}},\label{eq_fkpm}\\
 f_{K^0}=&\frac{\sigd+\sigs}{\sqrt{2}},\label{eq_fk0}
\end{align}
respectively.
For later convenience, we define $\sigma_q\equiv(\sigma_u+\sigma_d)/2$ and
$\delta\sigma_q\equiv(\sigma_d-\sigma_u)/2$.
In the isospin-symmetric limit,
$\sigu=\sigd=\sigq$ and $\left<\delta\sigma_q\right>=0$.
The parameters contained in the meson part of the Lagrangian,
$\mu^2$, $\lambda$, $\lambda'$, $B$, $A$, and the expectation values
of the sigma fields $\sigq$ and $\sigs$ in the free space are determined
so as to reproduce the observed meson masses and decay constants at $\rho=0$
in the isospin-symmetric limit;
%: The
explicit expressions
%of
for the meson
masses are presented in Appendix~\ref{app_in_med_mass}.
See Ref.~\cite{sakai2013} and references
%there in
therein
for more details.
%\sout{Here}
However,
% \sout{we note} 
it should be noted that some 
of the parameters in the meson
% \sout{part} 
sector in the present work are
different from those in Ref.~\cite{sakai2013} because of the
different choice of
%the
input parameters:
% \sout{of those used in this}
%In
in the present study, the parameters are determined 
so as to precisely reproduce
% \sout{the mass of} 
the $\eta$ meson mass among others.
% whose decay width into $3\pi$ we focus on in this paper 
%\sout{and we use the different input of the sigma-meson mass.}
% which is important for the decay width of the $\eta$ meson.
%\sout{In this calculation,}
Furthermore,
% \sout{we shall vary and treat the mass of}
the sigma meson mass in the free space 
%\sout{$m_\sigma(\rho=0)$} 
will be varied as an input parameter; we present the numerical
%result evaluated using $m_\sigma(\rho=0)=441,550,$
results evaluated using $m_\sigma(\rho=0)=441$, $550$,
and 668 MeV in Sect.~\ref{sec_result}.

The breaking of the isospin symmetry in the free space is taken into account
as a small perturbation, and thus we have
\begin{align}
 \dsq=\frac{(m_{K^0}^2-m_{K^\pm}^2)-(m_{\pi^0}^2-m_{\pi^\pm}^2)}{2B+2\lambda(2\sigq-\sigs)},
\end{align}
where Dashen's theorem \cite{dashen1969chiral} has been employed to omit
the leading-order correction of the pseudoscalar meson masses coming from
the electromagnetic effect.
%{\bf\sout{We show the input and determined parameters of
%the Lagrangian in Table \ref{tab_input2} and \ref{tab_output2}, and the
%obtained masses and mixing angles of mesons
%in the free space are presented in Table~\ref{tab_mesonmass2}.}}
 We show the input parameters except for the sigma meson mass
in Table~\ref{tab_input1}, and
the determined
% \sout{ones} 
values of the parameters in the Lagrangian  are given in
%Table~\ref{tab_input1} and
Tables~\ref{tab_output1}, \ref{tab_output2},
and \ref{tab_output3} for
$m_{\sigma}(\rho=0)=$441, 550, and 668 MeV, respectively.
The obtained meson masses and the mixing angles are presented in
Tables~\ref{tab_mesonmass1}, \ref{tab_mesonmass2}, and
\ref{tab_mesonmass3} for $m_\sigma(\rho=0)=441, 550,$ and 668 MeV,
respectively.
The definition of the mixing angle is given in
Appendix~\ref{app_in_med_mass}.
%{\bf\sout{Although we list the value 550 MeV as a typical value of $m_\sigma$ in
%the free space, we shall vary and treat it as an input parameter.}}
The parameters $\mu^2$,
$\lambda'$, the masses of $\sigma$, $f_0$, $a_0$, and the mixing
angles between them are changed along with $m_\sigma(\rho=0)$.
\begin{table}
 \centering
 \begin{tabular}[t]{c|c|c|c|c|c|c}
  $f_\pi$&$f_K$ &$m_\pi$ &$m_K$
  &$m_{\eta'}^2+m_{\eta}^2$&$m_q$&$m_{\sigma_0}$ \\\hline
  92.2&110.4 &138.04 &495.64 &$547.85^2+1092.0^2$&$(2.3+4.8)/2$&518.3,\
			  628,\ 775.2 \\
 \end{tabular}
 \caption{The input values of the basic observables used 
for the determination of the parameters in the Lagrangian.
% with $m_\sigma(\rho=0)=441$ MeV.
 The units of the meson decay constants and meson masses are MeV.
 We use the isospin average of the
 %PDG
 Particle Data Group value \cite{Agashe:2014kda} for
 $m_\pi,m_K,$ and $m_q$.
 The three values listed in the
 %column of
 $m_{\sigma_0}$ column are the input ones for
 $m_\sigma(\rho=0)=$ 441, 550, and 668 MeV.
 \label{tab_input1}}
\end{table}
\begin{table}
 \centering
 \begin{tabular}[t]{c|c|c|c|c|c|c|c|c|c|c|c}
  $\sigu$&$\sigd$ &$\sigs$&$m_u$ &$m_d$&$m_s$& $\mu^2$&$\lambda$ &$\lambda'$ &$A$ &$B$ &$g$ \\\hline 
%  MeV& MeV&MeV &MeV &MeV&MeV&MeV$^2$&-&-&MeV$^2$&MeV\\\hline
  64.8&65.6 &90.9 &2.12 &4.98 &106.1& 2.79$\times10^5$&45.6 &-1.70 &3.50$\times 10^5$ &4.67$\times10^3$ &6.61 \\
 \end{tabular}
 \caption{The values of the determined parameters in the Lagrangian with
 $m_\sigma(\rho=0)=441$ MeV.
 The units of the parameters $\mu^2$ and $A$
 %is
 are MeV$^2$,
 %that
 those of
 $\lambda$ and $\lambda'$ are dimensionless, and
 %that
 those of
 the
 others are MeV.
 \label{tab_output1}}
\end{table}
\begin{table}
 \centering
 \begin{tabular}[t]{c|c|c|c|c|c|c|c|c|c}
  $m_\sigma$&$m_{f_0}$ &$m_{a_0^0}$ &$m_{a_0^\pm}$&$m_{\eta'}$ &$m_\eta$ &$m_{\pi^0}$&$m_{\pi^\pm}$&$m_{K^\pm}$&$m_{K^0}$   \\ \hline
  441&1238 &1120 &1120&1093 &547.8 &137.9&138.0& 493.0&498.2 \\
 \end{tabular}
 \begin{tabular}[t]{c|c|c|c|c|c|c|c|c}
  $\theta_s$&$\theta_{a_0\sigma}$&$\theta_{a_0f_0}$&$\theta_{ps}$&$\theta_{\pi^0\eta}$&$\theta_{\pi^0\eta'}$&$\left<\bar{u}u\right>$& $\left<\bar{d}d\right>$ &
  $\left<\bar{s}s\right>$ \\
  \hline
  13.6&-0.40&5.5$\times10^{-2}$&-1.89&0.72&1.44$\times10^{-2}$&-283.0$^3$&-284.2$^3$ &-316.9$^3$
 \end{tabular}
 \caption{The obtained meson masses,
 %mixing angle of mesons,
 meson mixing angles,
 and quark
 condensate in the free space for the case of $m_\sigma(\rho=0)=441$ MeV.
 The units of the meson masses, quark condensates, and mixing angles are
 MeV, MeV$^3$, and
 %degree,
 degrees,
 respectively.
 \label{tab_mesonmass1}}
\end{table}

%\begin{table}
% \centering
% \begin{tabular}[t]{c|c|c|c|c|c|c}
%  $f_\pi$&$f_K$ &$m_\pi$ &$m_K$
%  &$m_{\eta'}^2+m_{\eta}^2$&$m_q$&$m_{\sigma_0}$ \\\hline
%  92.2&110.4 &138.04 &495.64 &$547.85^2+1092.0^2$&$(2.3+4.8)/2$&628 \\
% \end{tabular}
% \caption{The input values with $m_\sigma(\rho=0)=550$ MeV
% {\bf\sout{as an example}}.
% {\bf The details are the same as those in Table~\ref{tab_input1}.}
% \label{tab_input2}}
%\end{table}
\begin{table}
 \centering
 \begin{tabular}[t]{c|c|c|c|c|c|c|c|c|c|c|c}
  $\sigu$&$\sigd$ &$\sigs$&$m_u$ &$m_d$&$m_s$& $\mu^2$&$\lambda$ &$\lambda'$ &$A$ &$B$ &$g$ \\\hline 
%  MeV& MeV&MeV &MeV &MeV&MeV&MeV$^2$&-&-&MeV$^2$&MeV\\\hline
  64.8&65.6 &90.9 &2.12 &4.98 &106.1& 2.14$\times10^5$&45.6 &2.16
				  &3.50$\times 10^5$ &4.67$\times10^3$ &9.96 \\
 \end{tabular}
 \caption{The values of the determined parameters in the Lagrangian with
 $m_\sigma(\rho=0)=550$ MeV.
 The units of the parameters are the same as those in Table~\ref{tab_output1}.
 \label{tab_output2}}
\end{table}
\begin{table}
 \centering
 \begin{tabular}[t]{c|c|c|c|c|c|c|c|c|c}
  $m_\sigma$&$m_{f_0}$ &$m_{a_0^0}$ &$m_{a_0^\pm}$&$m_{\eta'}$ &$m_\eta$ &$m_{\pi^0}$&$m_{\pi^\pm}$&$m_{K^\pm}$&$m_{K^0}$   \\ \hline
  550&1247 &1120 &1120&1093 &547.8 &137.9&138.0& 493.0&498.2 \\
 \end{tabular}
 \begin{tabular}[t]{c|c|c|c|c|c|c|c|c}
  $\theta_s$&$\theta_{a_0\sigma}$&$\theta_{a_0f_0}$&$\theta_{ps}$&$\theta_{\pi^0\eta}$&$\theta_{\pi^0\eta'}$&$\left<\bar{u}u\right>$& $\left<\bar{d}d\right>$ &
  $\left<\bar{s}s\right>$ \\
  \hline
  15.7&-0.47&-4.66$\times 10^{-2}$&-1.89&0.72&1.44$\times10^{-2}$&-283.0$^3$&-284.2$^3$ &-316.9$^3$\\ 
 \end{tabular}
 \caption{The obtained meson masses,
 %mixing angle of mesons,
 meson mixing angles,
 and quark
 condensate in the free space for the case of $m_\sigma(\rho=0)=550$ MeV.
% {\bf\sout{as a typical value}}.
  The units of these values are the same as those in Table~\ref{tab_mesonmass1}.
 \label{tab_mesonmass2}}
\end{table}

%\begin{table}
% \centering
% \begin{tabular}[t]{c|c|c|c|c|c|c}
%  $f_\pi$&$f_K$ &$m_\pi$ &$m_K$
%  &$m_{\eta'}^2+m_{\eta}^2$&$m_q$&$m_{\sigma_0}$ \\\hline
%  92.2&110.4 &138.04 &495.64 &$547.85^2+1092.0^2$&$(2.3+4.8)/2$&755.15 \\
% \end{tabular}
% \caption{The input values with $m_\sigma(\rho=0)=668$ MeV.
% The details are same as those in Table~\ref{tab_input1}.
% \label{tab_input3}}
%\end{table}
\begin{table}
 \centering
 \begin{tabular}[t]{c|c|c|c|c|c|c|c|c|c|c|c}
  $\sigu$&$\sigd$ &$\sigs$&$m_u$ &$m_d$&$m_s$& $\mu^2$&$\lambda$ &$\lambda'$ &$A$ &$B$ &$g$ \\\hline 
  64.8&65.6 &90.9 &2.12 &4.98 &106.1&1.24$\times10^5$&45.6 &7.53
				  &3.50$\times 10^5$ &4.67$\times10^3$ &14.1 \\
 \end{tabular}
 \caption{The values of the determined parameters in the Lagrangian with
 $m_\sigma(\rho=0)=668$ MeV.
 The units of the parameters are same as those in Table~\ref{tab_output1}.
 \label{tab_output3}}
\end{table}
\begin{table}
 \centering
 \begin{tabular}[t]{c|c|c|c|c|c|c|c|c|c}
  $m_\sigma$&$m_{f_0}$ &$m_{a_0^0}$ &$m_{a_0^\pm}$&$m_{\eta'}$ &$m_\eta$ &$m_{\pi^0}$&$m_{\pi^\pm}$&$m_{K^\pm}$&$m_{K^0}$   \\ \hline
  668&1261 &1120 &1120&1093 &547.8 &137.9&138.0& 493.0&498.2 \\
 \end{tabular}
 \begin{tabular}[t]{c|c|c|c|c|c|c|c|c}
  $\theta_s$&$\theta_{a_0\sigma}$&$\theta_{a_0f_0}$&$\theta_{ps}$&$\theta_{\pi^0\eta}$&$\theta_{\pi^0\eta'}$&$\left<\bar{u}u\right>$& $\left<\bar{d}d\right>$ &
  $\left<\bar{s}s\right>$ \\
  \hline
  19.2&-0.6&-0.19&-1.89&0.72&1.44$\times10^{-2}$&-283.0$^3$&-284.2$^3$ &-316.9$^3$\\ 
 \end{tabular}
 \caption{The obtained meson masses,
 %mixing angle of mesons,
 meson mixing angles,
 and quark
 condensate in the free space for the case of $m_\sigma(\rho=0)=668$ MeV.
 The units of these values are the same as those in Table~\ref{tab_mesonmass1}.
 \label{tab_mesonmass3}}
\end{table}

%Denoting the
Writing
the
expectation values in the free space and the nuclear
medium
as
%by
$\left<\dots\right>_0$ and $\left<\dots\right>_\rho$,
respectively, we obtain the difference
%of
between the expectation values of
the scalar meson fields
$\delta\left<\sigma_i\right>=\left<\sigma_i\right>_\rho-\left<\sigma_i\right>_0\
(i=u,d,s)$
within the leading order
of the baryon density $\rho$ as follows:
%;
\begin{align}
 \begin{pmatrix}
  \delta\sigu\\
  \delta\sigd\\
  \delta\sigs
 \end{pmatrix}=-
 \begin{pmatrix}
  m_{uu}^2&m_{ud}^2&m_{us}^2\\
  m_{du}^2&m_{dd}^2&m_{ds}^2 \\
  m_{su}^2&m_{sd}^2&m_{ss}^2 
 \end{pmatrix}^{-1}
 \begin{pmatrix}
  g\rho_p\\
  g\rho_n\\
  0
 \end{pmatrix},
\end{align}
where $m_{ij}^2=\partial^2 V_{\rm eff}/\partial\sigma_i\partial\sigma_j$
($i,j=u,d,s$).
%The
Explicit expressions
%of
for
$m_{ij}^2$ are given in
 Appendix~\ref{app_in_med_mass}.
The
% \soutparameter} 
meson$-$nucleon coupling constant $g$ in Eq.~(\ref{eq_lsm_lagrangian})
%\sout{ which 
%is the parameter contained in the Lagrangian in the fermionic part and
%characterizes the strength of the coupling of
%the nucleon and mesons}
is determined so as to reproduce the 30\% reduction of
the quark condensate at the normal nuclear density
$\rho_0=0.17$ fm$^{-3}$, which is suggested from the analysis of the
deeply bound state of the pionic atom \cite{suzuki2004precision}.

\section{Medium effect on masses and vertices of mesons}
\label{sec_in_med_mass_and_vertex}
In this section, we evaluate the nuclear medium effects on the
self-energies and the vertices of the mesons. 
Our calculation is based on the expansion with respect to the Fermi
momentum $k_f$ of the nuclear medium;
we retain the terms up to $O(k_f^3)$ as mentioned in the previous
section.
The nucleon propagator in the nuclear medium $G(p;k_f)$ with $p$ being
the nucleon four-momentum reads 
\begin{align}
 iG(p;k_f)=&(\Slash{p}+m_N)\left\{\frac{i}{p^2-m_N^2+i\epsilon}-2\pi\delta(p^2-m_N^2)\theta(p_0)\theta(k_f-|\vec{p}|)\right\},
 \label{eq_nucleon_prop}
\end{align}
where the first and second terms are the
contributions from the nucleon propagation in the free space and the
nuclear hole state in the Fermi sea, respectively. 
%{\bf\sout{Using the Fermi momentum of proton (neutron) $k_f^{p(n)}$, its
%number density $\rho_{p(n)}$ is expressed as
%$\rho_{p(n)}=k_f^{p(n)3}/3\pi^2$.}}
At the leading order with respect to the Fermi momentum,
the meson loops do not have the nuclear
medium effect on the masses or the vertices of mesons.
%\sout{their contributions are renormalized into the free-space ones.}

The couplings of hadrons
$g_{\sigma_f\sigma_1\sigma_2},g_{\sigma_f\pi_1\pi_2},g_{\sigma_iN},$ and
$g_{\pi_iN}$ appearing in the following calculations are presented in
Appendix~\ref{app_meson_coupling}.

\subsection{Medium effect on meson masses}
\begin{figure}[t]
 \begin{center}
  \includegraphics[width=7cm]{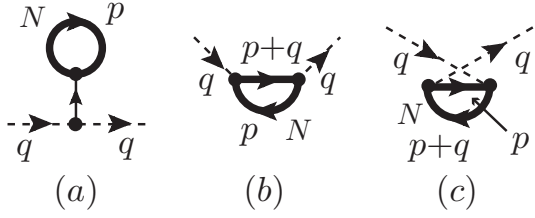}
  \caption{Diagrams contributing to the meson self-energies in the nuclear
  medium.
  The thin and thick lines represent the propagation of the scalar meson
  and the nucleon,
  %\sout{field}
  respectively.
  The external dashed lines represent the scalar or pseudoscalar mesons.}
  \label{diag_self_energy}
 \end{center}
\end{figure}

We evaluate the nucleon one-loop diagrams shown in
Fig.~\ref{diag_self_energy}
%\sout{contributing} 
%which give the leading contributions with respect to the Fermi momentum 
to the meson self-energies in the nuclear medium;
the leading contributions with respect to the Fermi momentum appear
from these diagrams.
%The
Diagram $(a)$ represents the contribution from the nucleon tadpole;
%, while the
diagrams $(b)$ and $(c)$
%from
the particle$-$hole excitations.
The self-energies of the scalar
and pseudoscalar mesons in the nuclear medium $\Sigma_s(\rho)$ and
$\Sigma_{ps}(\rho)$ are given as follows, respectively
(see Appendix~\ref{app_in_med_mass} for the details of the calculation):
\begin{align}
 -i\Sigma_s(\rho)=&-i\frac{g_{\sigma_kN}g_{\sigma_k\sigma_1\sigma_2}}{m_{\sigma_k}^2}\rho_{p(n)},\\
 -i\Sigma_{ps}(\rho)=&-i\frac{g_{\sigma_kN}g_{\sigma_k\pi_1\pi_2}}{m_{\sigma_k}^2}\rho_{p(n)}-i\frac{g_{\pi_1N}g_{\pi_2N}}{m_N}\rho_{p(n)}.\label{eq_self_ene_ps} 
\end{align}
We note that the particle$-$hole excitations
do not contribute to the self-energy of the scalar meson up to $O(k_f^3)$.

The sigma meson
%have
has
a finite decay width due to the coupling with the
$2\pi$ state.
We implement
%it
this
by the replacement
of the sigma propagator $i/(p^2-m_\sigma^2+i\epsilon)$ with
\begin{align}
 iG_\sigma(p^2)=\frac{i}{p^2-m_\sigma^2+i\Theta(p^2)},\label{eq_gfunc_sigma}
\end{align}
where $\Theta(p^2)$ denotes the width of the sigma meson. 
%In the
At
tree level, it reads
\begin{align}
 \Theta(p^2)=\frac{g_{\sigma\pi\pi}^2}{16\pi}\sqrt{1-4m_\pi^2/p^2}\theta(p^2-4m_\pi^2)
\equiv \Theta_{\rm tree}(p^2).\label{eq_gfunc_theta}
\end{align}
Various quantum effects would  modify the width from 
$\Theta_{\rm tree}(p^2)$ although its quantitative evaluation is beyond the scope of
the present work. In the present work, we shall use $\Theta_{\rm tree}(p^2)$
as a typical width and   
vary $\Theta(p^2)$ within 30\%
%per cent
around $\Theta_{\rm tree}(p^2)$ 
to see uncertainties of the results on  the $\eta\rightarrow3\pi$ decay width 
due to that of the width.
%\sout{where $\Theta(p^2)$ is the $\sigma\rightarrow2\pi$ decay width at the
%tree level.
%There is some possibility that the width of the sigma meson is
%modified by some quantum effect.
%The effect of the modification of the decay width of the sigma meson
%from the tree-level value given in Eq.~(\ref{eq_gfunc_theta}) on
%the $\eta\rightarrow3\pi$ decay width is discussed in
%Sect.~\ref{sec_numerical_results}.}

\subsection{Medium effect on vertices of mesons}
\begin{figure}[t]
 \begin{center}
  \includegraphics[width=7cm]{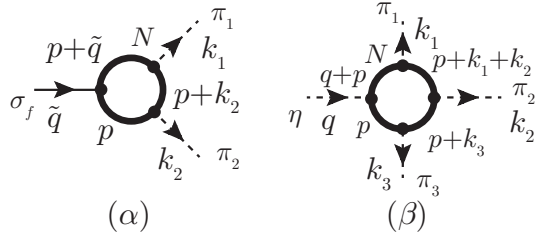}
  \caption{
  %The
  Diagrams contributing to the modification
  of
  the vertices of
  mesons.
  The thin and thick solid lines denote the propagation of the scalar
  meson and the nucleon, and the dashed line the propagation of the
  pseudoscalar meson.
  \label{fig_vertex_1loop}}
 \end{center}
\end{figure}

\begin{figure}[t]
 \centering
 \includegraphics[width=1.8cm]{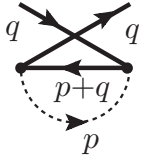}
 \caption{
 %The
 Diagram
 %which contribute
 contributing
 to the self-energy of the nucleon
 in the nuclear medium.}
 \label{diag_nucl_fock}
\end{figure}

The modifications of the vertices of mesons come from the diagrams shown
in Fig.~\ref{fig_vertex_1loop}.
Here, we
%denote
write
the momenta of the outgoing meson $\pi_i$,
the incoming scalar meson $\sigma_f$, and the $\eta$ meson in the
initial state
as
%by
$k_i$,
$\tilde{q}=k_1+k_2$, and $q=k_1+k_2+k_3$, respectively.
The leading corrections with respect to the Fermi momentum
$\Gamma_\alpha$ and $\Gamma_\beta$ of the scalar--2-pseudoscalar and the
4-pseudoscalar meson couplings from
%the
diagram $(\alpha)$ and $(\beta)$ in Fig.~\ref{fig_vertex_1loop} are
calculated to be
\begin{align}
 i\Gamma_\alpha(\rho)=&i\frac{g_{\sigma_fN}g_{\pi_1N}g_{\pi_2N}}{4m_N^2E_{\pi_1}E_{\pi_2}}(k_{\pi_1}\cdot
 k_{\pi_2})\rho_{p(n)},\label{eq_gamma_3pt_text}\\
 i\Gamma_\beta=&0,
\end{align}
respectively.
The calculational details of these are presented in
Appendix~\ref{app_meson_coupling}.
We use the nucleon mass in the free space in
Eq.~(\ref{eq_gamma_3pt_text}).
Its modification in the nuclear medium is
small by including the one-loop diagram given in
Fig.~\ref{diag_nucl_fock}
which gives the leading contribution from the nuclear medium.
The details of the calculation of the diagram
%is
are
given in
Appendix~\ref{app_nucl_mass}.

\section{Matrix element of $\eta\rightarrow 3\pi$ decay and numerical
 results\label{sec_result}}
\subsection{Matrix element of $\eta\rightarrow3\pi$ decay}
\begin{figure}[t]
 \begin{center}
  \includegraphics[width=15cm]{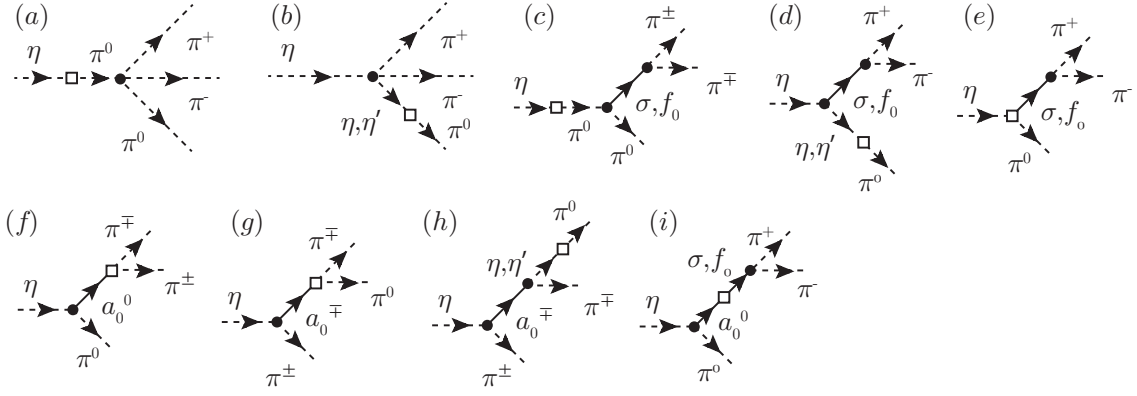}
  \caption{The tree-level diagrams contributing to the
  $\eta\rightarrow3\pi$ decay in the free space.
  The meanings of the lines are the same as those in
  Fig.~\ref{fig_vertex_1loop}.
  The vertex with the white box represents the effect of the isospin-symmetry
  breaking.
  \label{fig_tree_diag_decay}}
 \end{center}
\end{figure}

The tree-level diagrams that contribute to the $\eta\rightarrow3\pi$
decay in the free space are shown in Fig.~\ref{fig_tree_diag_decay}.
We divide the contributions into three parts:
the first part, $\mathcal{M}_{\rm contact}^\ppp$, represented by
%the
diagrams $(a)$ and $(b)$, is the contributions from the 4-pseudoscalar
meson contact vertices.
The second one, $\mathcal{M}_{\rm isoscalar}^\ppp$, given by
%the
diagrams $(c)$, $(d)$, $(e)$, and $(i)$, has the isoscalar mesons
$\sigma$ and $f_0$ in the intermediate state.
The last one, $\mathcal{M}_{\rm isovector}^\ppp$, obtained from
%the
diagrams $(f)$, $(g)$, $(h)$, and $(i)$, contains the isovector meson
$a_0$.
We shall see that $\mathcal{M}_{\rm isoscalar}^\ppp$ gives the most
significant contribution
to the in-medium $\eta\rightarrow3\pi$ decay width, which will be
discussed in the next section.
%The
Diagrams $(a)$, $(b)$, $(d)$, and $(h)$ are the contributions from
the $\eta-\pi_3$ and $\eta'-\pi_3$ mixing by ISB.
The $\sigma-\sigma_3$ and $f_0-\sigma_3$ mixings from ISB
provide the contribution shown in
%the
diagram $(i)$ in
Fig.~\ref{fig_tree_diag_decay}.

The matrix element of the $\eta\rightarrow\ppp$ decay in the free space is
written as follows:
%;
\begin{align}
 \mathcal{M}^{\rm tree}_{\eta\rightarrow\ppp}=&\mathcal{M}_{\rm
 contact}^\ppp+\mathcal{M}_{\rm isoscalar}^\ppp+\mathcal{M}_{\rm
 isovector}^\ppp,\\
 \mathcal{M}_{\rm
 contact}^\ppp=&2(-\sin\theta_{\pi^0\eta})g_{\pi_3\pi_3\pi^+\pi^-}+(-\sin\theta_{\pi^0\eta'}\frac{\sin 2\theta_{ps}}{2}+2\sin\theta_{\pi^0\eta}\sin^2\theta_{ps})g_{\eta_0\eta_0\pi^+\pi^-}\notag\\
&+(\sin\theta_{\pi^0\eta'}\cos^2
 \theta_{ps}-2\sin\theta_{\pi^0\eta}\sin
 2\theta_{ps})g_{\eta_0\eta_8\pi^+\pi^-}\notag\\
&+(\sin\theta_{\pi^0\eta}\frac{\sin
 2\theta_{ps}}{2}-2\sin\theta_{\pi^0\eta}\cos^2\theta_{ps})g_{\eta_0\eta_0\pi^+\pi^-},\\
  \mathcal{M}_{\rm
 isoscalar}^\ppp=&-\left\{g_{\sigma\eta\pi^0}\frac{1}{s-m_\sigma^2}g_{\sigma\pi^+\pi^-}+g_{\eta\pi^0f_0}\frac{1}{s-m_{f_0}^2}g_{f_0\pi^+\pi^-}\right.\notag\\
&+2(-\sin\tpe)g_{\sigma\pi_3\pi_3}\frac{1}{s-m_\sigma^2}g_{\sigma\pi^+\pi^-}+2(-\sin\tpe)g_{\pi_3\pi_3f_0}\frac{1}{s-m_{f_0}^2}g_{f_0\pi^+\pi^-}\notag\\
 &+2g_{\sigma\eta\eta}(\sin\tpe)\frac{1}{s-m_{\sigma}^2}g_{\sigma\pi^+\pi^-}+2g_{f_0\eta\eta}(\sin\tpe)\frac{1}{s-m_{f_0}^2}g_{f_0\pi^+\pi^-}\notag\\ 
&+g_{\eta\eta'\sigma}(\sin\tpep)\frac{1}{s-m_{\sigma}^2}g_{\sigma\pi^+\pi^-}+g_{\eta\eta'
 f_0}(\sin\tpep)\frac{1}{s-m_{f_0}^2}g_{f_0\pi^+\pi^-}\notag\\
 &\left.+g_{a_0\pi^0\eta}\frac{1}{s-m_\sigma^2}
 g_{\sigma\pi^+\pi^-}(-\sin\tss)+\frac{1}{s-m_{f_0}^2}g_{f_0\pi^+\pi^-}(-\sin\tsf)\right\},\label{eq_melem_ppp_tree_scalar}\\
  \mathcal{M}_{\rm
 isovector}^\ppp=&-\left\{g_{\eta\pi^-a_0^+}\frac{1}{t-m_{a_0^-}^2}g_{a_0^-\pi^+\pi^0}+g_{\eta\pi^+a_0^-}\frac{1}{u-m_{a_0^+}^2}g_{a_0^+\pi^-\pi^0}+g_{\eta\pi^0a_0^0}\frac{1}{s-m_{a_0^0}^2}g_{a_0^0\pi^+\pi^-}\right.\notag\\
 &+g_{a_0^0\pi^0\eta}\frac{1}{s-m_{a_0^0}^2}g_{\sigma\pi^+\pi^-}(\sin\tss)+g_{a_0^2\pi^0\eta}\frac{1}{s-m_{a_0^0}^2}g_{f_0\pi^+\pi^-}(\sin\tsf)\notag\\
 &+g_{\eta\pi^-a_0^+}\frac{1}{t-m_{a_0^-}^2}g_{\eta\pi^+a_0^-}(\sin\tpe)+g_{\eta\pi^+a_0^-}\frac{1}{u-m_{a_0^+}^2}g_{\eta\pi^-a_0^+}(\sin\tpe)\notag\\
&\left.+g_{\eta\pi^-a_0^+}\frac{1}{t-m_{a_0^-}^2}g_{\eta'\pi^+a_0^-}(\sin\tpep)+g_{\eta\pi^+a_0^-}\frac{1}{u-m_{a_0^+}^2}g_{\eta'\pi^-a_0^+}(\sin\tpep)\right\}.
\end{align}
The Mandelstam variables $s,t$, and $u$ are defined as
$s=(p_\eta-p_{\pi^0})^2$, $t=(p_\eta-p_{\pi^+})$, and
$u=(p_\eta-p_{\pi^-})$, respectively.
The couplings of mesons in the matrix elements are presented
in Appendix~\ref{app_tree_vertex}.

\begin{figure}[t]
 \begin{center}
  \includegraphics[width=10cm]{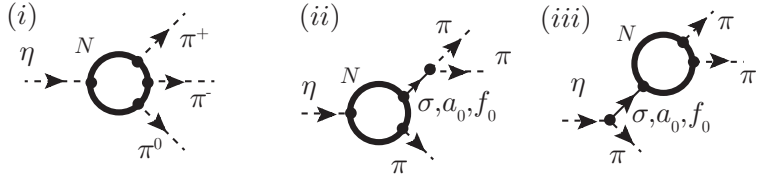}
  \caption{Diagrams contributing to the $\eta\rightarrow3\pi$ decay in
  the nuclear medium.
  The meanings of the lines are
  %the
  identical to those in Fig.~\ref{fig_vertex_1loop}.
  \label{fig_one_loop_diag_decay}}
 \end{center}
\end{figure}

The one-loop diagrams for the $\eta\rightarrow\ppp$ decay
in
the nuclear medium are shown in Fig.~\ref{fig_one_loop_diag_decay}.
%\com{the leading contributions of the Fermi momentum are obtained from
%these diagrams}.
%The
Diagram $(i)$ in Fig.~\ref{fig_one_loop_diag_decay} leads to the
in-medium correction
of the 4-pseudoscalar meson coupling,
while
%the
diagrams $(ii)$ and $(iii)$ in Fig.~\ref{fig_one_loop_diag_decay}
lead to
that of the scalar--2-pseudoscalar meson coupling.
The resultant modification of the $\eta\rightarrow\ppp$ decay amplitude
$\delta\mathcal{M}^{\rm loop}_{\eta\rightarrow\ppp}$ is given as follows:
\begin{align}
 &\delta\mathcal{M}^{\rm loop}_{\eta\rightarrow\ppp}=\delta\mathcal{M}^\ppp_{\rm
 mixing}+\delta\mathcal{M}^\ppp_{\rm
 isoscalar}+\delta\mathcal{M}^\ppp_{\rm isovector}(\rho),\\
 &\delta\mathcal{M}^\ppp_{\rm
 mixing}=+2(-\sin\tpe)\Gamma_{\pi_3\pi_3\pi^+\pi^-}+2(\sin\tpe)\Gamma_{\eta\eta\pi^+\pi^-}+(\sin\tpep)\Gamma_{\eta\eta'\pi^+\pi^-},
 \label{eq_deltamelem_contact}\\
 &\delta\mathcal{M}^\ppp_{\rm isoscalar}=-\frac{\Gamma_{\sigma\eta\pi_3}g^0_{\sigma\pi^+\pi^-}}{s-m_\sigma^2}-\frac{g^0_{\sigma\eta\pi_3}\Gamma_{\sigma\pi^+\pi^-}}{s-m_\sigma^2}-\frac{\Gamma_{f_0\eta\pi_3}g^0_{f_0\pi^+\pi^-}}{s-m_{f_0}^2}-\frac{g^0_{f_0\eta\pi_3}\Gamma_{f_0\pi^+\pi^-}}{s-m_{f_0}^2}\notag\\
 &-2(-\sin\tpe)\left\{\frac{\Gamma_{\sigma\pi_3\pi_3}g^0_{\sigma\sigma\pi^+\pi^-}}{s-m_\sigma^2}+\frac{g^0_{\sigma\pi_3\pi_3}\Gamma_{\sigma\sigma\pi^+\pi^-}}{s-m_\sigma^2}+\frac{\Gamma_{f_0\pi_3\pi_3}g^0_{f_0\sigma\pi^+\pi^-}}{s-m_{f_0}^2}+\frac{g^0_{f_0\pi_3\pi_3}\Gamma_{f_0\sigma\pi^+\pi^-}}{s-m_{f_0}^2}\right\}\notag \\
 &-2(\sin\tpe)\left\{\frac{\Gamma_{\sigma\eta\eta}g^0_{\sigma\pi^+\pi^-}}{s-m_\sigma^2}+\frac{g^0_{\sigma\eta\eta}\Gamma_{\sigma\pi^+\pi^-}}{s-m_\sigma^2}+\frac{\Gamma_{f_0\eta\eta}g^0_{f_0\pi^+\pi^-}}{s-m_{f_0}^2}+\frac{g^0_{f_0\eta\eta}\Gamma_{f_0\pi^+\pi^-}}{s-m_{f_0}^2}\right\} \notag\\
 &-(\sin\tpep)\left\{\frac{\Gamma_{\sigma\eta\eta'}g^0_{\sigma\pi^+\pi^-}}{s-m_\sigma^2}+\frac{g^0_{\sigma\eta\eta'}\Gamma_{\sigma\pi^+\pi^-}}{s-m_\sigma^2}+\frac{\Gamma_{f_0\eta\eta'}g^0_{f_0\pi^+\pi^-}}{s-m_{f_0}^2}+\frac{g^0_{f_0\eta\eta'}\Gamma_{f_0\pi^+\pi^-}}{s-m_{f_0}^2}\right\} \notag\\
 &-(\sin\theta_{\sigma_3\sigma})\left\{\frac{\Gamma_{a_0^0\eta\pi_3}g^0_{\sigma\pi^+\pi^-}}{s-m_\sigma^2}+\frac{g^0_{a_0^0\eta\pi_3}\Gamma_{\sigma\pi^+\pi^-}}{s-m_\sigma^2}\right\}-(\sin\theta_{\sigma_3f_0})\left\{\frac{\Gamma_{a_0^0\eta\pi_3}g^0_{f_0\pi^+\pi^-}}{s-m_{f_0}^2}+\frac{g^0_{a_0^0\eta\pi_3}\Gamma_{f_0\pi^+\pi^-}}{s-m_{f_0}^2}\right\}, \label{eq_melem_ppp_loop_scalar}\\
 &\delta\mathcal{M}^\ppp_{\rm isovector}=-\frac{\Gamma_{a_0^+\eta\pi^-}g^0_{a_0^-\pi_3\pi^+}}{t-m_{a_0^\pm}^2}-\frac{\Gamma_{a_0^-\eta\pi^+}g^0_{a_0^+\pi_3\pi^-}}{u-m_{a_0^\pm}^2}-\frac{\Gamma_{a_0^0\eta\pi_3}g^0_{a_0^0\pi^+\pi^-}}{s-m_{a_0^0}^2}\notag\\
 &-(-\sin\theta_{\sigma_3\sigma})\left\{\frac{\Gamma_{a_0^0\eta\pi_3}g^0_{\sigma\pi^+\pi^-}}{s-m_{a_0^0}^2}+\frac{g^0_{a_0^0\eta\pi_3}\Gamma_{\sigma\pi^+\pi^-}}{s-m_{a_0^0}^2}\right\}-(-\sin\theta_{\sigma_3f_0})\left\{\frac{\Gamma_{a_0^0\eta\pi_3}g^0_{f_0\pi^+\pi^-}}{s-m_{a_0^0}^2}+\frac{g^0_{a_0^0\eta\pi_3}\Gamma_{f_0\pi^+\pi^-}}{s-m_{a_0^0}^2}\right\}\notag\\
 &-(\sin\tpe)\left\{\frac{\Gamma_{a_0^+\eta\pi^-}g^0_{a_0^-\eta\pi^+}}{t-m_{a_0^-}^2}+\frac{g^0_{a_0^+\eta\pi^-}\Gamma_{a_0^-\eta\pi^+}}{t-m_{a_0^-}^2}+\frac{\Gamma_{a_0^-\eta\pi^+}g^0_{a_0^+\eta\pi^-}}{u-m_{a_0^-}^2}+\frac{g^0_{a_0^-\eta\pi^+}\Gamma_{a_0^+\eta\pi^-}}{u-m_{a_0^-}^2}\right\}\notag \\
 &-(\sin\tpep)\left\{\frac{\Gamma_{a_0^+\eta\pi^-}g^0_{a_0^-\eta'\pi^+}}{t-m_{a_0^-}^2}+\frac{g^0_{a_0^+\eta\pi^-}\Gamma_{a_0^-\eta'\pi^+}}{t-m_{a_0^-}^2}+\frac{\Gamma_{a_0^-\eta\pi^+}g^0_{a_0^+\eta'\pi^-}}{u-m_{a_0^-}^2}+\frac{g^0_{a_0^-\eta\pi^+}\Gamma_{a_0^+\eta'\pi^-}}{u-m_{a_0^-}^2}\right\}.
\end{align}
The vertices with the superscript $0$ are the couplings of the mesons in
the free space.
The one-loop vertex correction in the nuclear medium
$\Gamma_{\sigma_f\pi_i\pi_j}$ is given in Eq.~(\ref{eq_gamma_3pt_text}).
The couplings of mesons and meson--nucleon are shown in
Appendices~\ref{app_tree_vertex} and \ref{app_meson_nucl}.
Here, the scalar meson masses and the mixing angles are the in-medium
values.
The contribution from $\delta\mathcal{M}_{\rm mixing}^\ppp$ in
Eq.~(\ref{eq_deltamelem_contact}) does not exist because the
correction of
the 4-pseudoscalar meson vertex $\Gamma_\beta$ vanishes as mentioned
before.
The matrix element in the nuclear medium is given as
$\mathcal{M}_{\eta\rightarrow\pi^+\pi^-\pi^0}(\rho)=\mathcal{M}^{\rm
tree}_{\eta\rightarrow\pi^+\pi^-\pi^0}+\delta\mathcal{M}^{\rm
loop}_{\eta\rightarrow\pi^+\pi^-\pi^0}$.

Denoting the matrix element of the $\eta\rightarrow\ppp$ process by
$\mathcal{M}_{\eta\rightarrow\ppp}(s,t,u)$, the decay
amplitude of the $\eta\rightarrow3\pi^0$ decay
$\mathcal{M}_{\eta\rightarrow3\pi^0}(s,t,u)$ is expressed as
\begin{align}
 \mathcal{M}_{\eta\rightarrow3\pi^0}(s,t,u)=\mathcal{M}_{\eta\rightarrow\ppp}(s,t,u)+\mathcal{M}_{\eta\rightarrow\ppp}(t,u,s)+\mathcal{M}_{\eta\rightarrow\ppp}(u,s,t),
 \label{eq_melem_eta3p0}
\end{align}
owing to the Bose symmetry of the mesons \cite{bijnens2007eta}; note that
our calculation only includes the leading-order effects of ISB.

\subsection{Numerical results}
\label{sec_numerical_results}
The decay width $\Gamma$ of the $\eta$ meson to three pions is evaluated
with the integration over the three-body phase space as
\begin{align}
 \Gamma=&\frac{1}{(2\pi)^3}\frac{1}{32m_\eta^3}\frac{1}{n!}\int ds\int
 dt\left|\mathcal{M}\right|^2, 
\end{align}
where $n$ is the number of
%the
identical particles, and
the phase space is evaluated using the meson masses in the free space.
In the present study, the mass of the sigma meson in the free space
$m_\sigma(\rho=0)$ is treated as a varying input as mentioned in
Sect.~\ref{sec_setup}:
We show the results of the calculations using
$m_\sigma(\rho=0)=441,550$, and $668$ MeV as typical values in the
following.
The widths of the sigma meson in the free space for the respective sigma
masses are evaluated
%to be
as 124, 296, and 605 MeV.

The $\eta\rightarrow3\pi$ decay width in the free space is fairly well
reproduced in our model; 
the $\eta\rightarrow\ppp$ and $3\pi^0$ decay widths in the free space are
obtained as about 200 and 290 eV, which are about 70\% of the respective
experimental values \cite{Agashe:2014kda}.
The discrepancy may be attributed to the insufficient treatment of FSI,
as mentioned in Sect.~\ref{intro}.

Here, we should remark on the valid range of the density of our calculation.
Our calculation only takes account of the leading-order contribution of
$k_f$, so the validity is limited in the small
baryon density, as mentioned in Sect.~\ref{sec_in_med_mass_and_vertex}.
It is also problematic that the mass of the sigma meson
%become
becomes less than
$2m_\pi$ at $\rho=0.11$, $0.14$, and $0.16$ fm$^{-3}$ for
$m_\sigma(\rho=0)=441$, $550$, and $668$ MeV, respectively,
which
%may
could show that the tree-level approximation of the sigma meson
%would be
is inadequate due to its small mass.
Therefore, the valid density region of our calculation would be lower than
the density although it may depend on the mass of the sigma meson in
the free space.

\begin{figure}[t]
 \begin{center}
  \begin{minipage}[t]{0.45\hsize}
   \begin{center}
   \includegraphics[width=6.5cm]{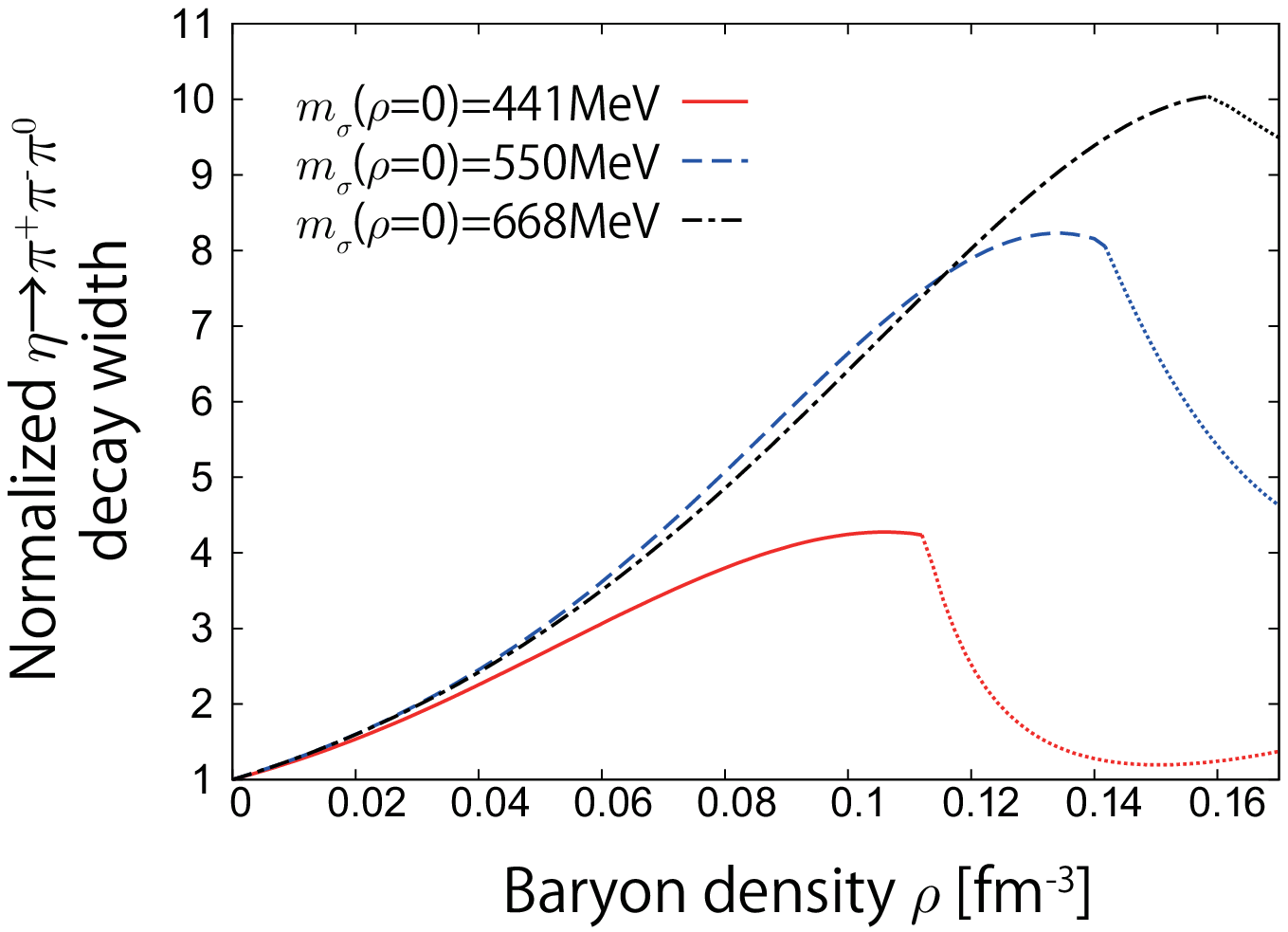}
   \end{center}
  \end{minipage}
  \begin{minipage}[t]{0.45\hsize}
   \begin{center}
   \includegraphics[width=6.5cm]{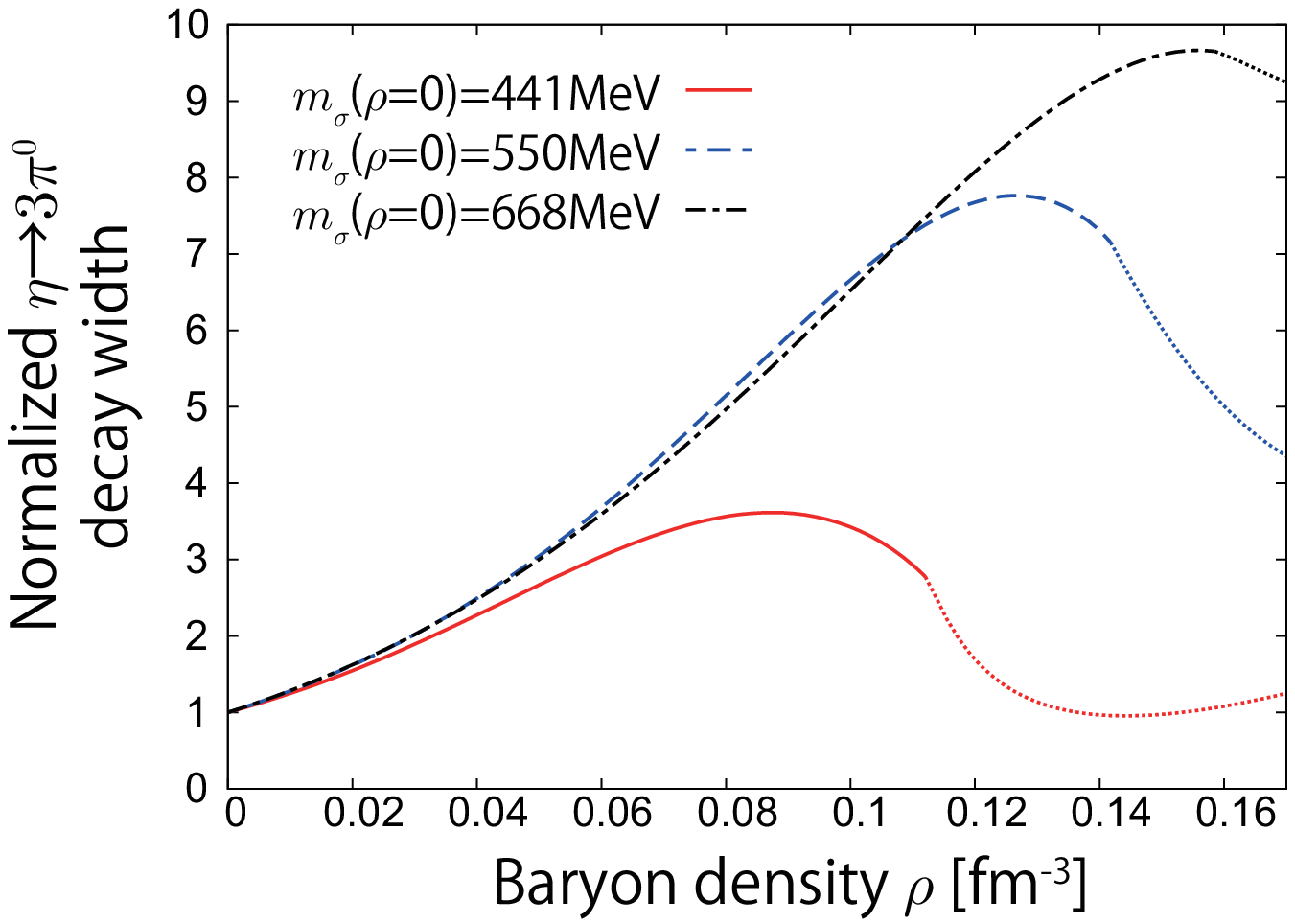}
   \end{center}
  \end{minipage}
  \caption{
  %The
  Plots of the decay widths of
% {\bf\sout{the}}
  $\eta\rightarrow\ppp$ 
  (left) and $\eta\rightarrow3\pi^0$ (right) in the symmetric nuclear medium.
  The red solid, blue dashed, and black dashed-dotted lines are the plot
  of the decay
  widths with $m_\sigma(\rho=0)=$441, 550, and 668 MeV.
  A dotted line is used when the mass of the sigma meson
  is less than $2m_\pi$ for all inputs of $m_\sigma(\rho=0)$.
  The decay widths of the $\eta\rightarrow\ppp$ process at $\rho=0$ are
  192, 182, and 220 eV with $m_\sigma(\rho=0)=$441, 550, and 668 MeV.
  Those of $\eta\rightarrow3\pi^0$ are 279, 265, and 317 eV.
  \label{fig_in_medium_width}}
 \end{center}
\end{figure}

\begin{figure}[t]
 \centering
 \begin{minipage}[t]{0.3\hsize}
  \begin{center}
   \includegraphics[width=4.5cm]{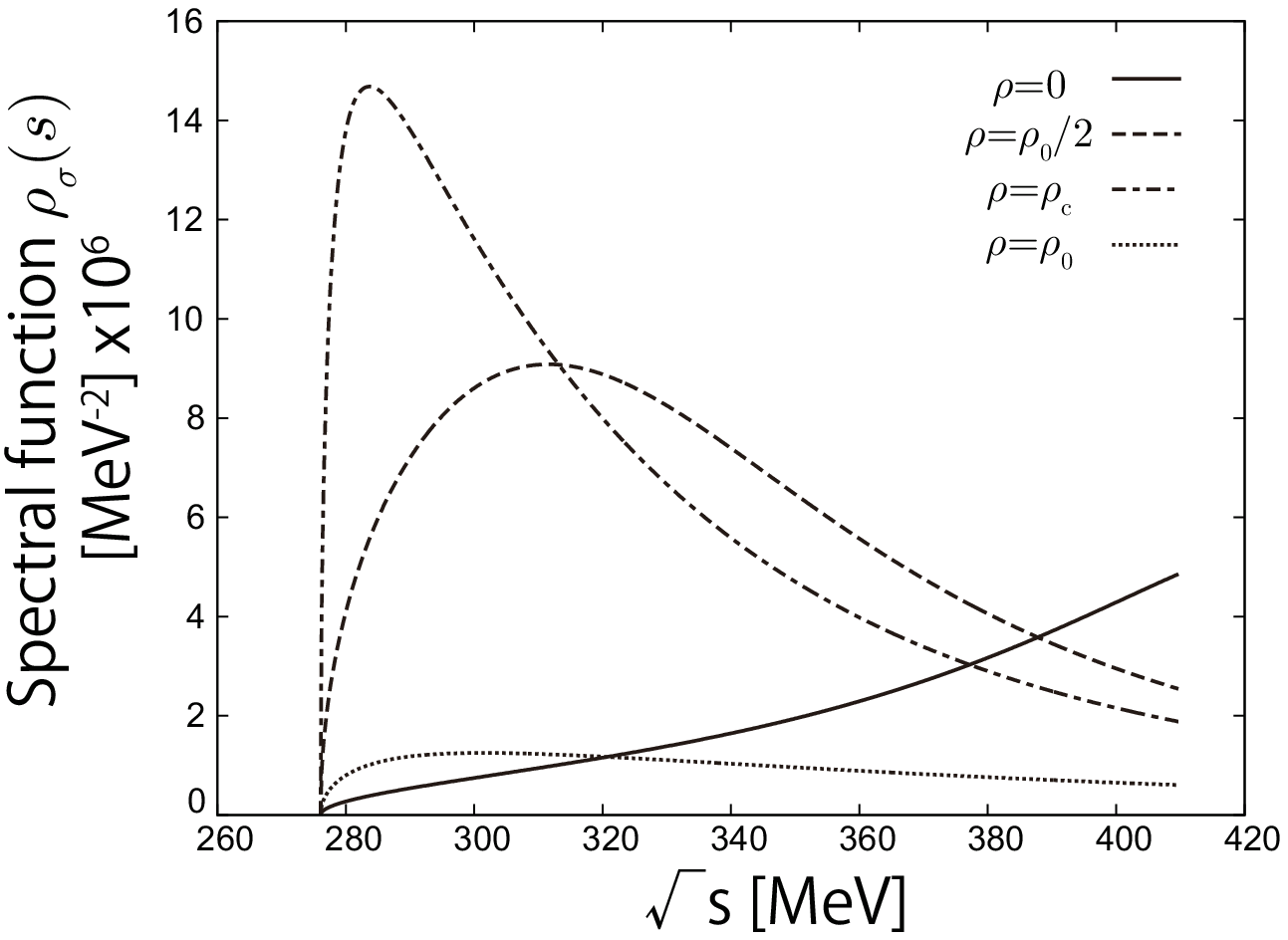}
  \end{center}
 \end{minipage}
  \begin{minipage}[t]{0.3\hsize}
   \begin{center}
    \includegraphics[width=4.5cm]{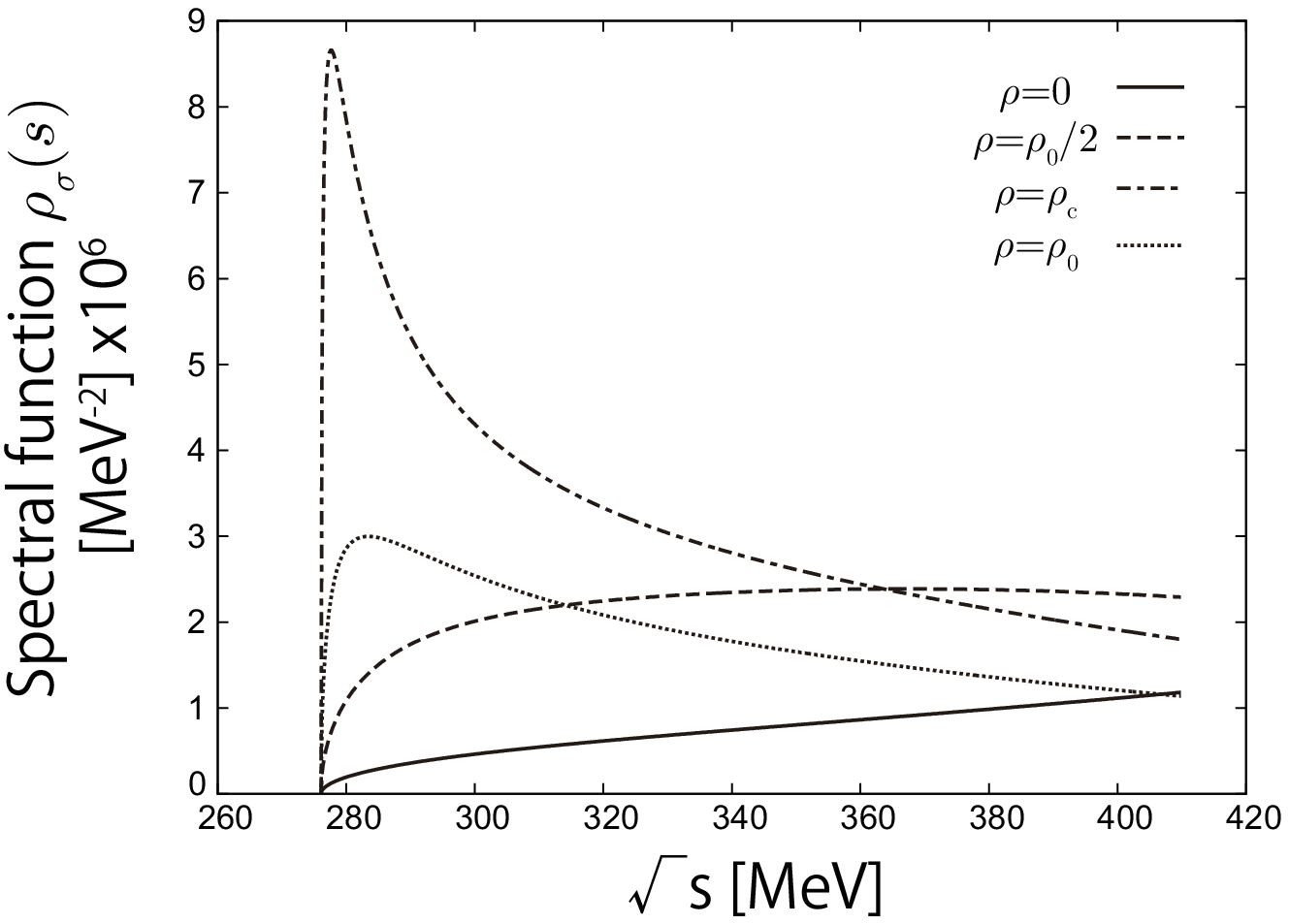}
   \end{center}
  \end{minipage}
  \begin{minipage}[t]{5cm}
   \begin{center}
    \includegraphics[width=4.5cm]{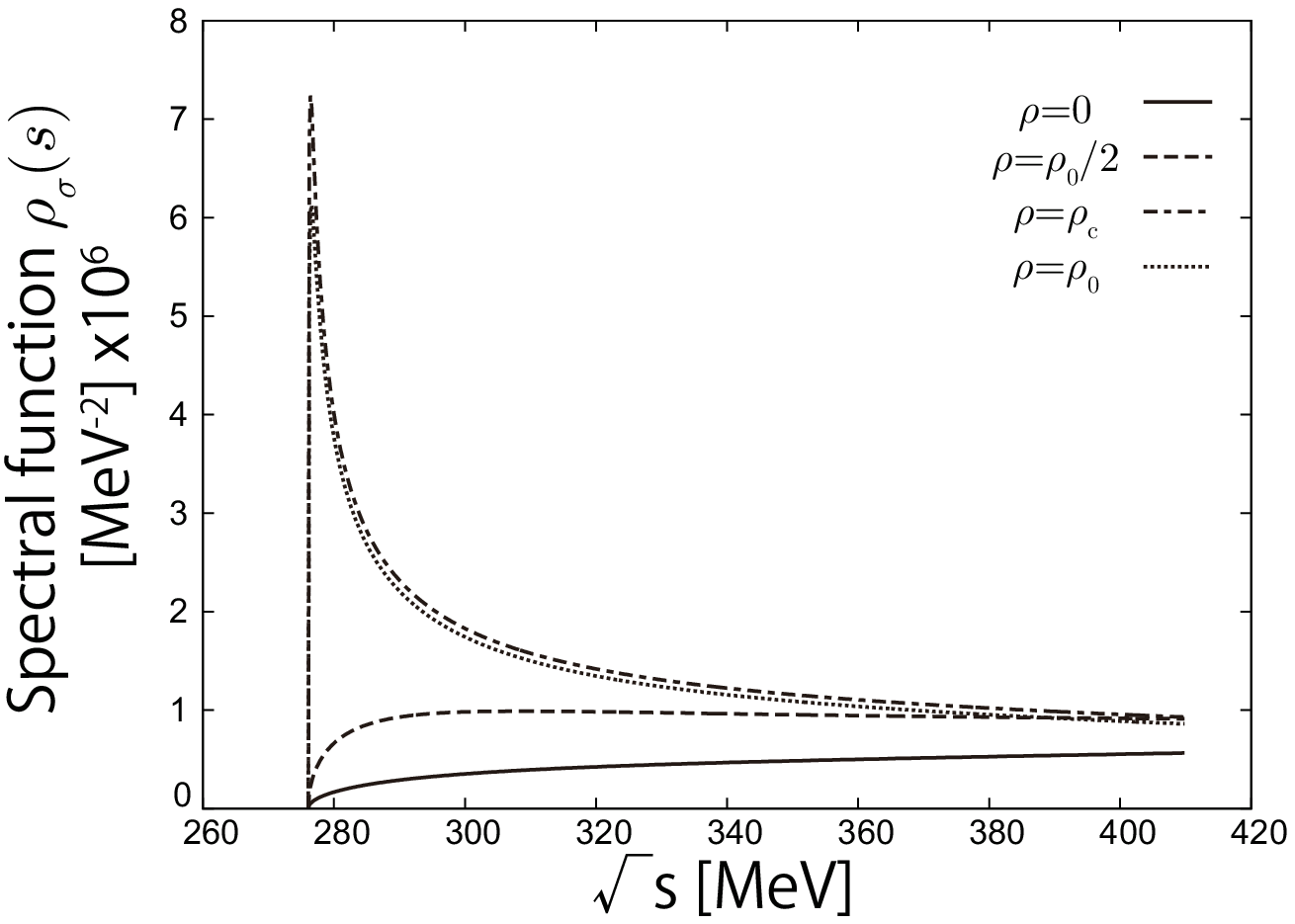}
   \end{center}
  \end{minipage}
 \caption{The spectral functions of the sigma meson $\rho_\sigma(s)$ with
 the mass of the sigma meson in the free space 441 MeV (left), 550 MeV
 (center), and 668 MeV (right) varying the baryon number density $\rho$.
 The horizontal axis is $\sqrt{s}$ and the vertical axis is the spectral
 function times $10^6$.
 The plot of the spectral function against $\sqrt{s}$ is limited in the
 range between $\sqrt{s_{\rm max}}=m_\eta-m_{\pi^0}$ and $\sqrt{s_{\rm
 min}}=2m_{\pi^\pm}$.
 The spectral functions at $\rho=0$, $\rho_0/2$, $\rho_c$, and $\rho_0$
 are plotted in each figure.
 Here, $\rho_c=0.1$, 0.13, and 0.15 fm$^{-3}$ for
 $m_\sigma(\rho=0)=441$, 550, and 668 MeV where the
 $\eta\rightarrow3\pi$ decay width is most enhanced for
 %the
 each
 $m_\sigma(\rho=0)$.
 \label{fig_spectral_sigma}}
\end{figure}

In Fig.~\ref{fig_in_medium_width}, we show the plot of the density
dependence of the decay
%width
widths
$\Gamma_{\eta\rightarrow\ppp}(\rho)$ and
$\Gamma_{\eta\rightarrow3\pi^0}(\rho)$ in the symmetric 
nuclear medium up to $\rho_0$ with varying $m_\sigma(\rho=0)$.
In the figures,
%the
dotted lines are used when the mass of the sigma
meson becomes less than $2m_\pi$.
From the figure, one can see an enhancement of the
decay width
%along with the increase of the
with increasing
baryon density $\rho$.
We mention that both of the decay widths of $\eta\rightarrow\ppp$ and
$3\pi^0$ show a peak structure in Fig.~\ref{fig_in_medium_width}.
Here, we write $\rho_c$ as the density at which the decay width is
most enhanced for each $m_\sigma(\rho=0)$.
$\rho_c$
%are
is
given as $0.1$, 0.13, and 0.15 fm$^{-3}$ for
$m_\sigma(\rho=0)=$441, 550, and 668 MeV, respectively.
The amount of the enhancement
at $\rho=\rho_c$
depends on
$m_\sigma(\rho=0)$;
%at $\rho=\rho_c$;
it becomes at most about four to ten times larger than the value in the free
space.
This originates from the stronger coupling of the sigma meson with the
two-pion state for the larger
$m_{\sigma}(\rho=0)$.
On the other hand, the dependence on $m_\sigma(\rho=0)$ is
relatively small at
%the
%density
densities lower than $\rho_0/2$.
The appearance of the peak at $\rho=\rho_c$ in
Fig.~\ref{fig_in_medium_width} is associated with the approach of the
sigma mass to the $2\pi$ threshold.
To see this, we show the spectral function of the sigma meson
$\rho_\sigma(s)=-\im G_\sigma(s)/\pi$ in
Fig.~\ref{fig_spectral_sigma}, where $G_\sigma(s)$ is the Green function
of the sigma meson given in Eq.~(\ref{eq_gfunc_sigma}).
One can see a growth of the peak near the $2\pi$ threshold accompanied by the
increase of the baryon density in Fig.~\ref{fig_spectral_sigma}.
The enhancement of the spectral function is most significant at
$\rho=\rho_c$.
Thus, the contributions from the sigma meson propagation
in
Eqs.~(\ref{eq_melem_ppp_tree_scalar}) and
(\ref{eq_melem_ppp_loop_scalar}) are enhanced in the nuclear medium.
The locations of the cusps appearing in Fig.~\ref{fig_in_medium_width}
correspond to the densities at which the mass of the sigma meson
becomes as small as $2m_\pi$.

\begin{figure}[t]
 \begin{center}
  \includegraphics[width=6.5cm]{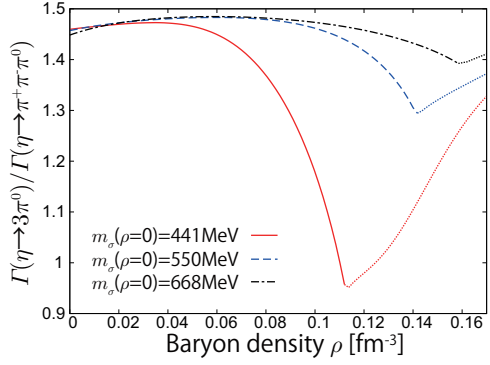}
  \caption{The density dependence of the ratio of the decay width
  of $\eta\rightarrow3\pi^0$ to that of $\eta\rightarrow\ppp$ in the nuclear
  medium.
  The lines
  %represent
  have
  the same meanings as those in Fig.~\ref{fig_in_medium_width}.
  \label{fig_in_medium_width_ratio}}
 \end{center}
\end{figure}

Next, we show the density dependence of the ratio of
$\Gamma_{\eta\rightarrow3\pi^0}(\rho)$ to
$\Gamma_{\eta\rightarrow\ppp}(\rho)$ in
Fig.~\ref{fig_in_medium_width_ratio}.
The plot shows a decrease of the ratio around $\rho_c$ compared with the
value at $\rho=0$; 
i.e. the enhancement of the $\eta\rightarrow3\pi^0$ decay width is
weaker than that of the $\eta\rightarrow\ppp$ decay around
$\rho=\rho_c$.
In the $\eta\rightarrow3\pi^0$ decay,
we simply write the contribution from the sigma meson
to $\eta\rightarrow3\pi^0$ $\mathcal{M}^{3\pi^0}_{\rm sigma}$ as 
\begin{align}
 \mathcal{M}^{3\pi^0}_{\rm
 sigma}=&-\frac{g_{\sigma\eta\pi}g_{\sigma\pi\pi}}{s-m_{\sigma}^2}-\frac{g_{\sigma\eta\pi}g_{\sigma\pi\pi}}{t-m_\sigma^2}-\frac{g_{\sigma\eta\pi}g_{\sigma\pi\pi}}{u-m_\sigma^2}.
 \label{eq_melem_3p0_scalar}
\end{align}
When the Mandelstam variable $s$ is near the $2\pi$ threshold,
$s\sim 4m_\pi^2\equiv s_{\rm min}$ and $t,u\sim (m_\eta^2-m_\pi^2)/2\equiv
t_{\rm max}$. 
When the mass of the sigma meson is reduced within the range between
$s_{\rm min}$ and $t_{\rm max}$ as the result of the softening of the
sigma meson in the nuclear medium, the second and third terms of
Eq.~(\ref{eq_melem_3p0_scalar}) have
%the
opposite signs to the first
term.
Thus, the
%way
degree
of the enhancement of the $\eta\rightarrow3\pi^0$ decay width is
%suppressed
small
compared with the $\eta\rightarrow\ppp$ decay when
the mass of the sigma meson decreases in the nuclear medium.
This suppression becomes more severe in the case of the smaller
$m_\sigma(\rho=0)$ because the sigma-pole contribution
plays a major role.

\begin{figure}
 \begin{minipage}[t]{0.49\hsize}
  \includegraphics[width=7.2cm]{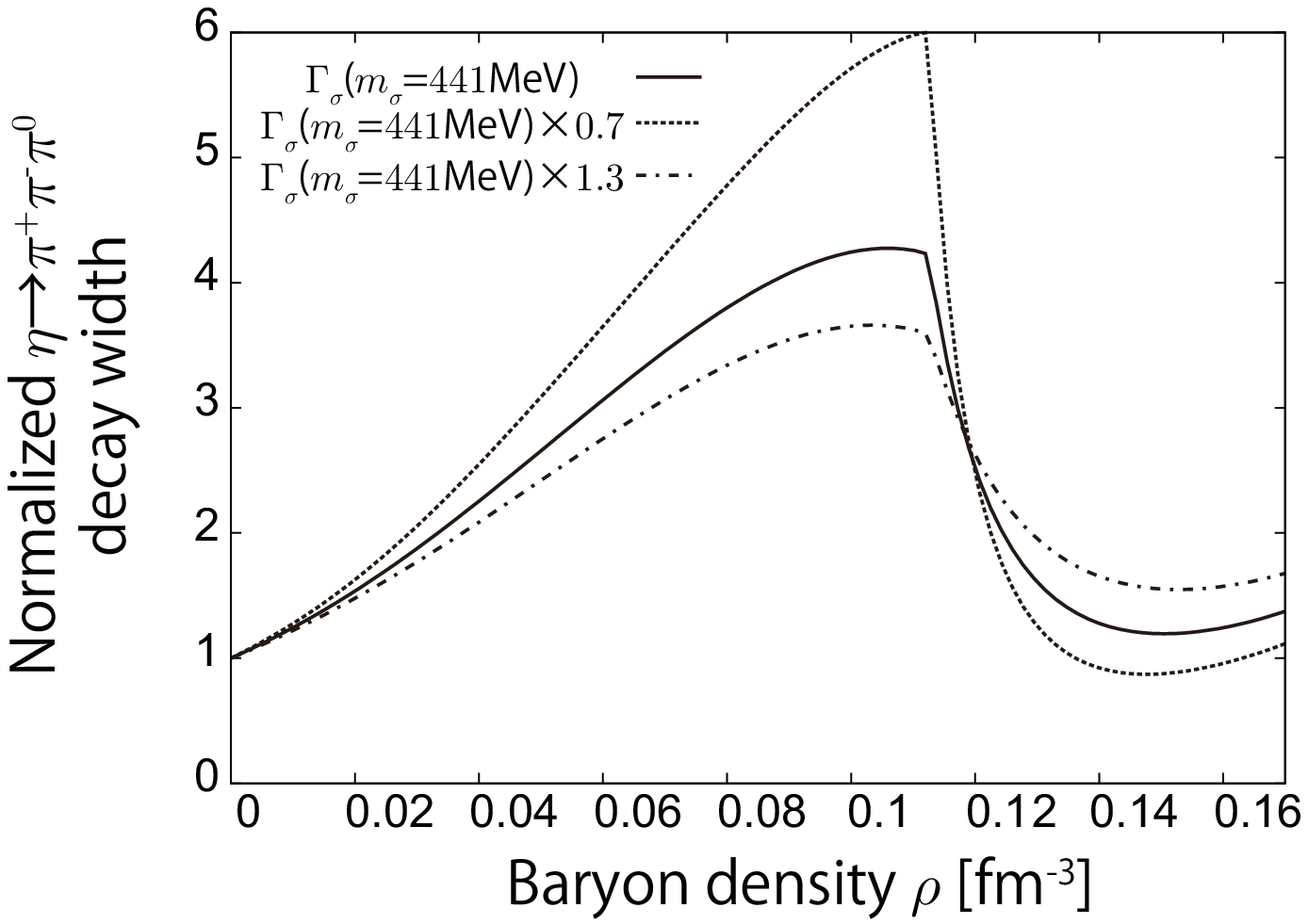}
 \end{minipage}
 \hfill
 \begin{minipage}[t]{0.49\hsize}
  \includegraphics[width=7.2cm]{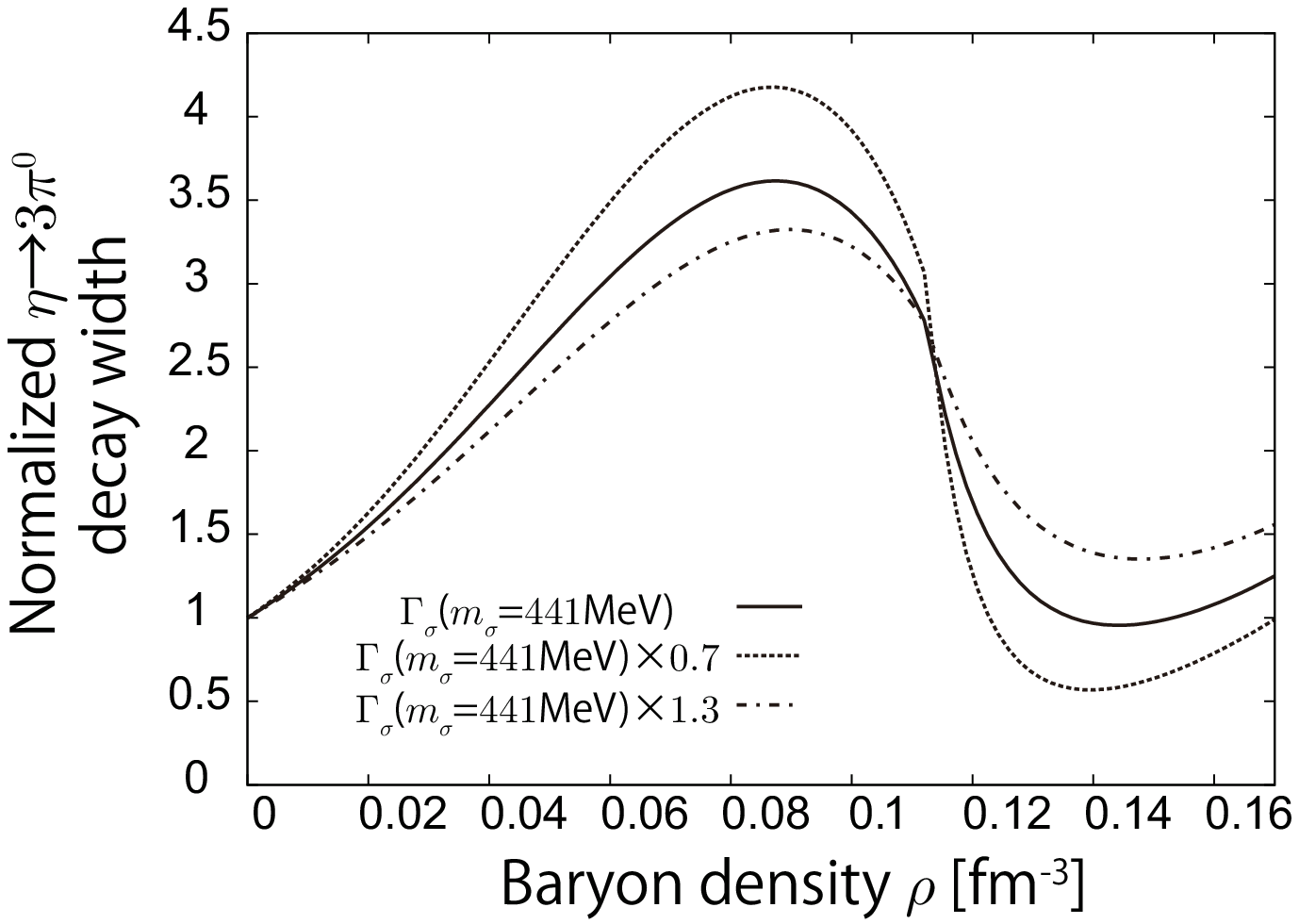}
 \end{minipage}
  %\end{figure}
%\begin{figure}[t]
 \begin{minipage}[t]{0.49\hsize}
  \includegraphics[width=7.2cm]{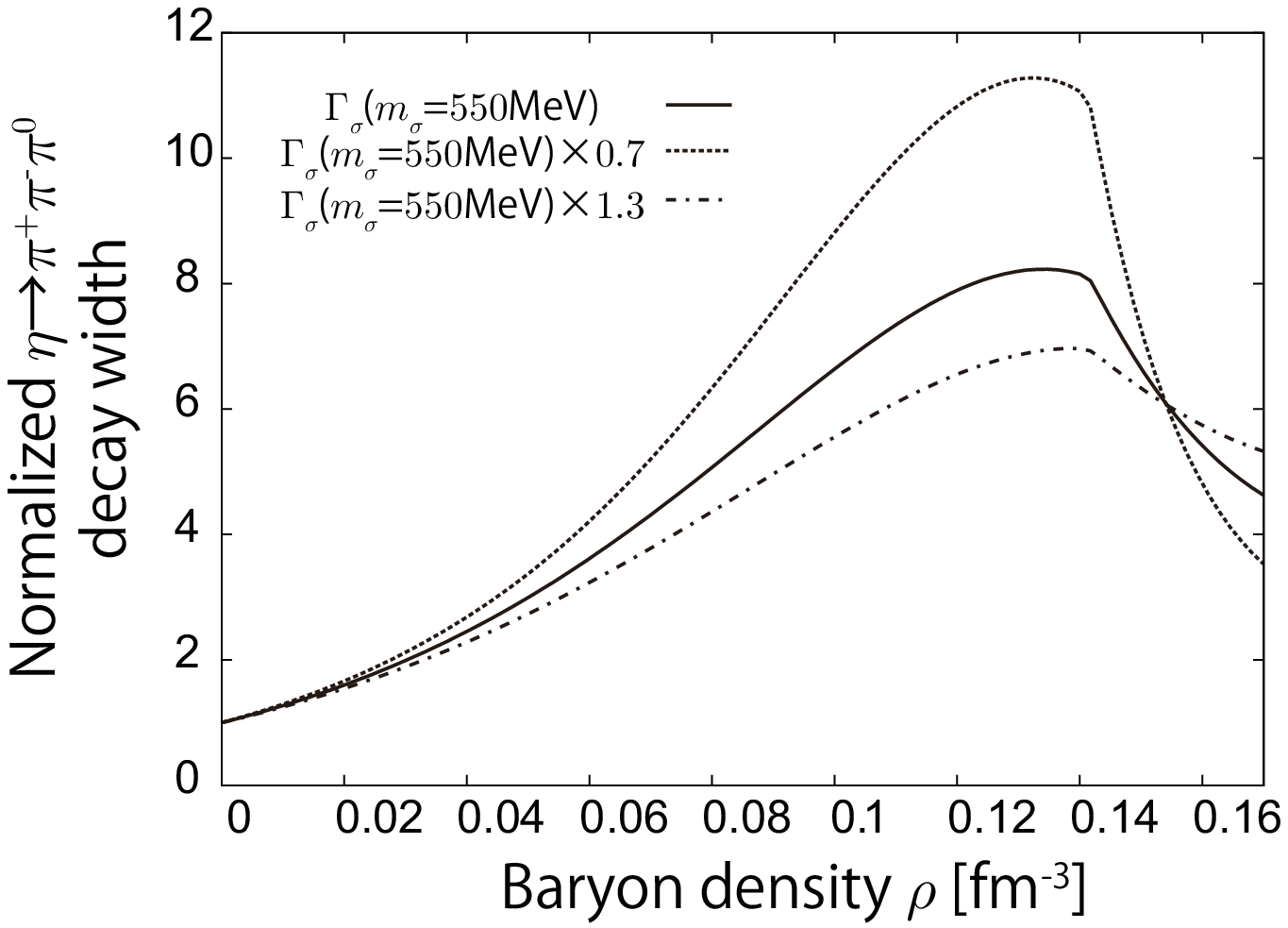}
 \end{minipage}
 \hfill
 \begin{minipage}[t]{0.49\hsize}
  \includegraphics[width=7.2cm]{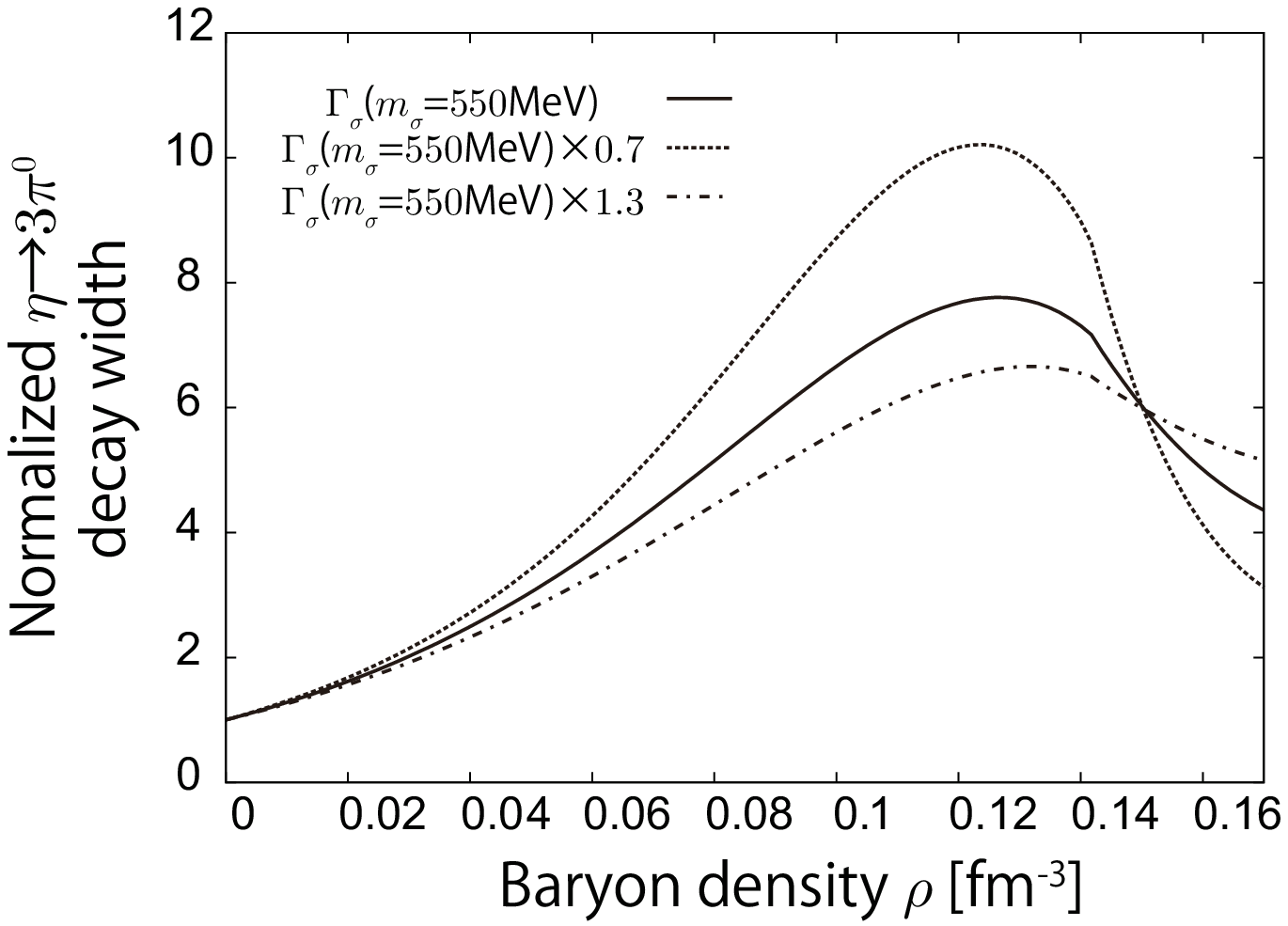}
 \end{minipage}
% \end{figure}
% \begin{figure}[t]
  \begin{minipage}[t]{0.49\hsize}
   \includegraphics[width=7.2cm]{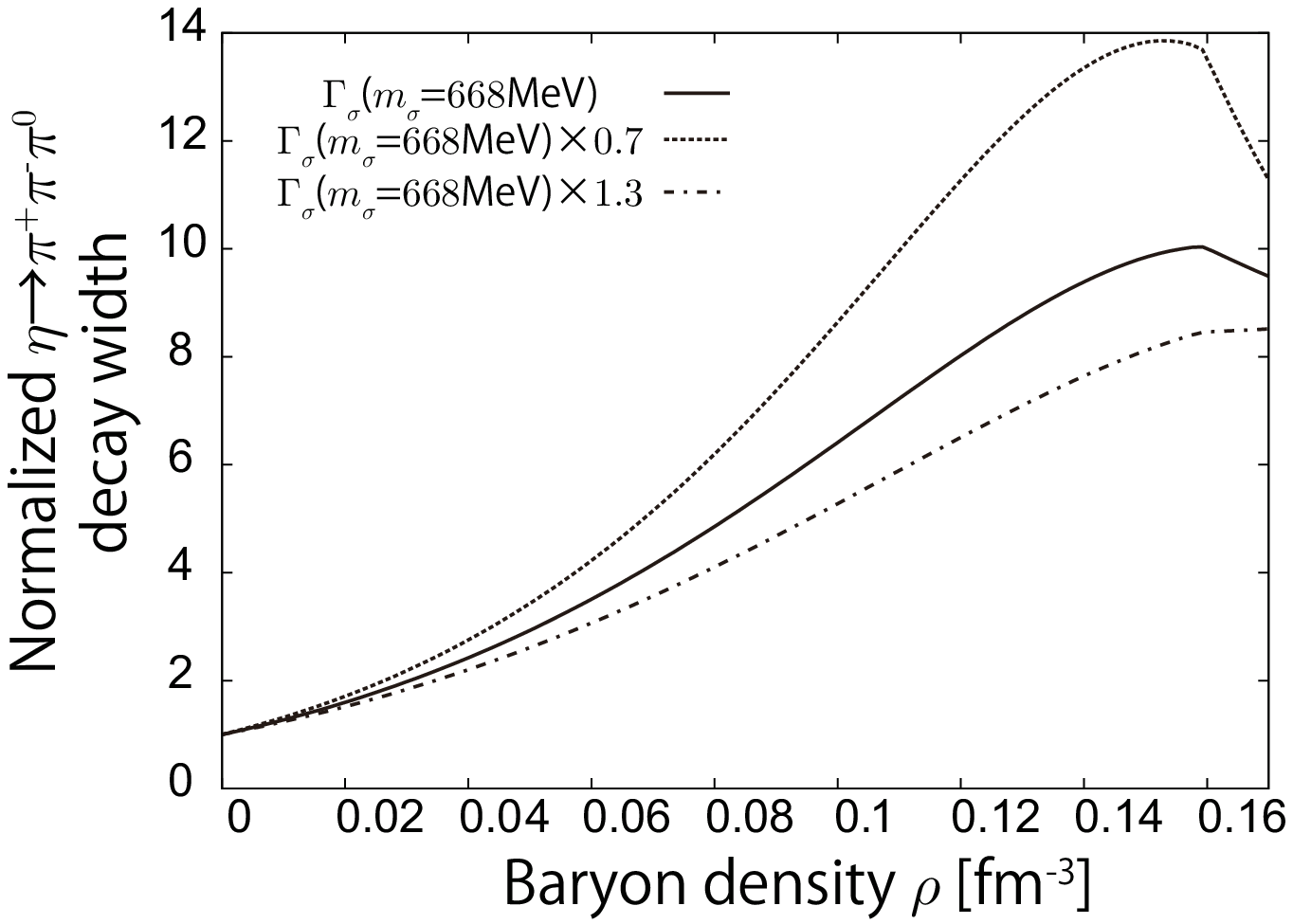}
  \end{minipage}
  \hfill
  \begin{minipage}[t]{0.49\hsize}
   \includegraphics[width=7.2cm]{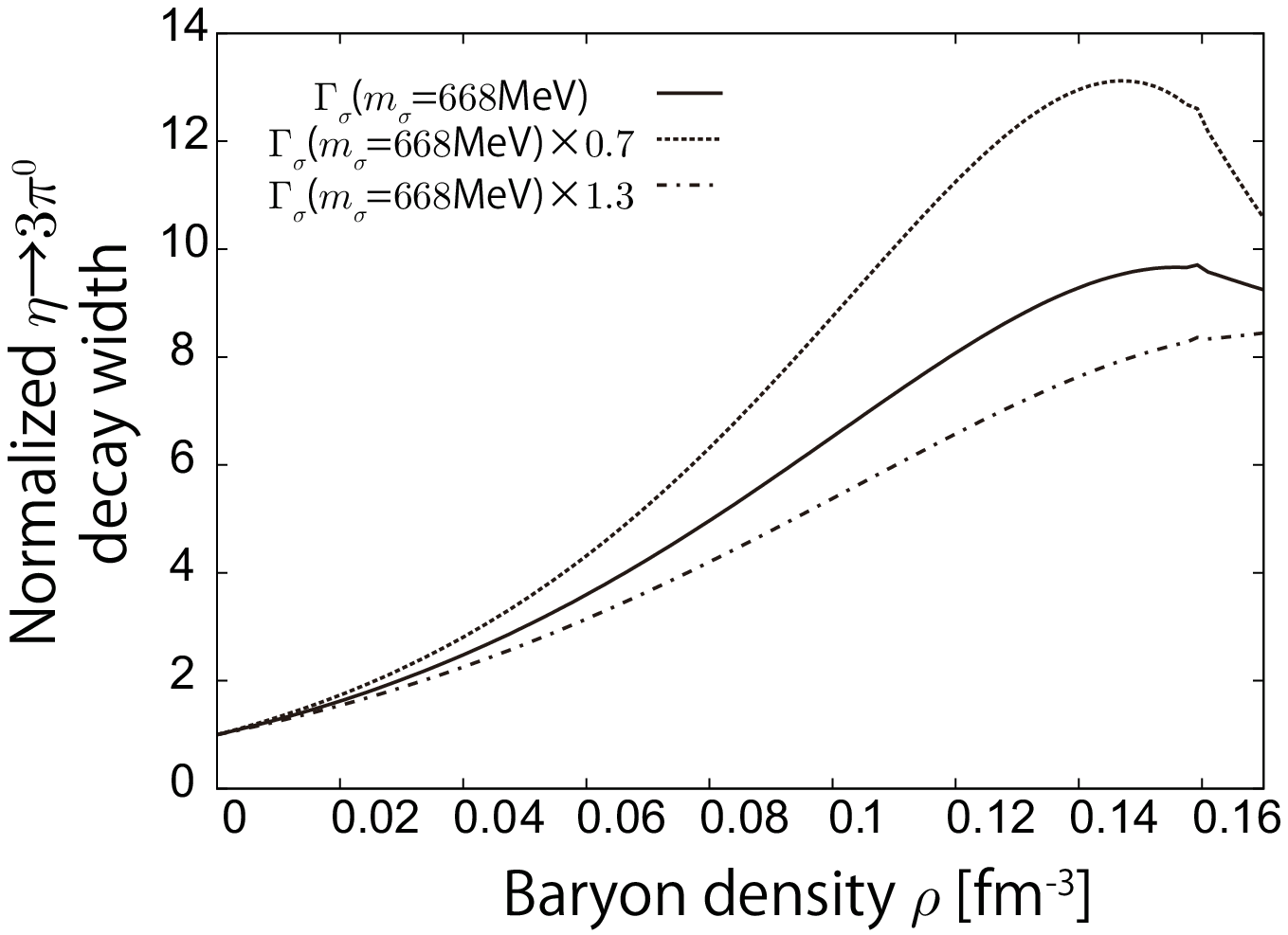}
  \end{minipage}
  \caption{
 %The
 Plot of the decay widths of the $\eta\rightarrow\ppp$
 (left panels) and
 $3\pi^0$ (right panels) processes by including the
 %thirty-percent
 30\%
 uncertainty
 of the decay width of the sigma meson.
 The upper, middle, and lower panels for each column are
 %the plot
 plots
 of the decay width using $m_\sigma(\rho=0)=441,550,$ and $668$ MeV,
 respectively.
 The
 % width of the sigma meson at the tree level
 $\eta\rightarrow3\pi$ decay width evaluated using the tree-level decay
 width of the sigma meson
 is plotted with
 %the
 a
 solid line and
 that evaluated using
 the width of the sigma meson multiplied by a factor 0.7
 (1.3)
 %for the plot of the
 is plotted with a
 dotted (dash-dotted) line.
 The decay widths of $\eppp$ at $\rho=0$ evaluated with the
 decay width of the sigma meson multiplied by 0.7 (1.3) are 197 (185),
 178 (184), and 199 (235) eV for $m_\sigma(\rho=0)=441,550,$ and 668
 MeV, respectively.
 Those of $\eta\rightarrow3\pi^0$ are 286 (271), 260 (269), and 288
 (340) eV.}
 \label{fig_width_mod}
\end{figure}

We comment on the uncertainty of the above results
% \sout{owing} 
due to that of the decay width of the sigma meson;
%First, we mention that of the decay width of the sigma meson.
it can be modified from the tree-level value by
% \sout{the} 
quantum effects as we mentioned in
Sect.~\ref{sec_in_med_mass_and_vertex}.
In Fig.~\ref{fig_width_mod}, we show the plots of the decay widths of
%the
$\eta\rightarrow\ppp$ and
$3\pi^0$ in the nuclear medium taking account of the
%thirty-percent
30\%
modification
of the width of the sigma meson from the tree-level value;
%Figure
Fig.~\ref{fig_width_mod} is plotted using $\Theta_{\rm
tree}(p^2)\times0.7$ and
$\times1.3$ as the decay width of the sigma
meson where $\Theta_{\rm tree}(p^2)$ is the decay width at the tree level
given in Eq.~(\ref{eq_gfunc_theta}).
As one can see from these figures, the difference 
is
% \sout{particularly} 
large
in the high-density region. 
The difference from the tree-level value is about 40\% at most, which is
significant with the large mass of the sigma meson in the free space.

\section{Conclusions and concluding remarks\label{sec_conclusion}}
In the present work, we have investigated the decay widths of the
$\eta\rightarrow\ppp$ and $3\pi^0$ processes in the nuclear medium using
the linear sigma model focusing on the role of the softening of the
sigma meson;
the softening of the sigma meson in the
nuclear medium is naturally taken into account in the model.

In the free space, our model
gives
%the
a
fairly good agreement with the experimental value;
the obtained decay width is about 70\% of the observed one
\cite{Agashe:2014kda}.
The deficiency of our result in the free space may be attributed to an
incomplete treatment of FSI, such as neglect of the
$\rho$ meson contribution.

For the calculation in the nuclear medium, the validity of our
calculation is restricted within the small Fermi momentum $k_f$ of the
nuclear medium
due to our limited calculations including only the leading-order
contributions of $k_f$.

Our calculation shows that the enhancement
%become
becomes
prominent when the
sigma mass is near the $2\pi$ threshold. 
The decay widths become at most about four to ten times larger than the
value at the free space, depending on the mass of the sigma meson in
the free space which is a parameter of our model.
The mechanism of the
enhancement is attributed to the reduction of the mass of the sigma meson 
associated with the chiral restoration.
It is worth emphasizing that
the enhancement of the decay width is significant
even at small density:
For example, the decay width at $\rho=\rho_0/2$ is several times larger
than that in the free space, and does not
%so much
depend
so much
on the mass of the sigma meson in the free space.
Thus, our claim is that the $\eta\rightarrow3\pi$ decay in the
nuclear medium
%can be a new possible
could be a possible new
probe for the chiral restoration.

It is worth mentioning
%, here,
here
that the density dependence of the decay width of
$\eta\rightarrow3\pi^0$ is smaller than that of $\eppp$, owing to
the Bose symmetry of the identical $3\pi^0$ in the
final state and the softening of the sigma meson in the nuclear medium.
The difference
%of
between
the density dependences of the two decay widths
should have
%a
significant importance in
%a
the
possible experimental confirmation of 
the findings of the present study because 
the experimental determination of the ratio of the two final states 
seems much easier than that of the respective widths,
%that
which
are small.

There
%are
is
considerable room for
%the
elaboration of the present calculation.
 The contribution from the $\rho$ meson can modify the density
dependence of the decay width of the $\eta$ meson.
However, the modification is expected to be small at low density 
on which the present work is focused, although
%\sout{particularly at the large}
it can be significant at high density.
%\sout{ though the effect would be small in the low density
%where we focused on in this study.}
We should
%make
undertake a
more complete treatment of FSI and the nuclear medium.
In particular,
%the
inclusion of the higher-order contribution of $k_f$ is
needed for
%the
a
realistic argument;
%it
this
contains the effect of the Fermi sea, which is necessary
to determine the ground state in the nuclear medium properly.
These are, however, beyond the scope of the present work, and 
%left as
are left for
future projects.

\section*{Acknowledgments}
We are grateful to  K. Imai and H. Fujioka for informative discussions 
on the issues
%for
of
a possible experimental confirmation of the present
study.
S.S. is a JSPS fellow and appreciates the support of a JSPS
Grant-in-Aid (No.~25-1879). 
The work by T.K. is partially supported by
%the
a
Grant-in-Aid for
Scientific Research from JSPS (No.~24340054).

\appendix
\section{Calculation of meson self-energy in the nuclear medium}
\label{app_in_med_mass}
In this appendix, we perform the calculation of the meson self-energies in
the nuclear medium and show the meson masses appearing in our
calculations.
The diagrams contributing to the meson masses are shown in
Fig.~\ref{diag_self_energy}.

First, we evaluate
%the
diagram $(a)$ in Fig.~\ref{diag_self_energy}.
The contribution to the scalar meson self-energy is evaluated as follows;
\begin{align}
 -i\Sigma^{(a)}=&-\frac{ig_{\sigma_f\sigma_1\sigma_2}g_{\sigma_f N}}{-m_{\sigma_f}^2}\tr\int\frac{d^4p}{(2\pi)^4}(-2\pi)(\Slash{p}+m_N)\delta(p^2-m_N^2)\theta(p_0)\theta(k_f-|\vec{p}|)\notag\\
 \sim&-i\frac{g_{\sigma_f\sigma_1\sigma_2}g_{\sigma_fN}}{m_{\sigma_f}^2}\frac{4m_N}{(2\pi)^3}\frac{1}{2m_N}\frac{4\pi}{3}k_f^{p(n)3}=-i\frac{g_{\sigma_f\sigma_1\sigma_2}g_{\sigma_fN}}{m_{\sigma_f}^2}\rho_{p(n)}. 
\end{align}
Here, $k_f^{p(n)}$ is the Fermi momentum of
the
proton (neutron).
From the first to the second
%lines
line, we have made an approximation
$E_N(\vec{p})\sim m_N$, assuming that the nucleon mass is large enough.
The pseudoscalar meson self-energy from
%the
diagram $(a)$ in
Fig.~\ref{diag_self_energy} is calculated in much the same way as the case of
the scalar meson one,
other than the coupling
$g_{\sigma_f\sigma_1\sigma_2}$ and $g_{\sigma_f\pi_1\pi_2}$.

Next, we evaluate the contributions from
%the
diagrams $(b)$ and $(c)$ in
Fig.~\ref{diag_self_energy} for the scalar and pseudoscalar mesons.
%The
Diagram $(b)$ for the self-energy of the scalar meson is given as
\begin{align}
 -i\Sigma_s^{(b)}=&-\tr\int\frac{d^4p}{(2\pi)^4}(-ig_{\sigma_1N})\frac{i(\Slash{p}+\Slash{q}+m_N)}{(p+q)^2-m_N^2}(-ig_{\sigma_2N})(-2\pi)(\Slash{p}+m_N)\delta(p^2-m_N^2)\theta(p_0)\theta(k_f-|\vec{p}|)\notag\\
 =&-ig_{\sigma_1N}g_{\sigma_2N}\int\frac{d^4p}{(2\pi)^3}\frac{4[p\cdot(p+q)+m_N^2]}{2p\cdot
 q+q^2}\frac{1}{2E_N(\vec{p})}\delta(p_0-E_N(\vec{p}))\theta(k_f-|\vec{p}|)\notag\\
 \sim&-4ig_{\sigma_1N}g_{\sigma_2N}\frac{2m_N^2+m_Nm_s}{2m_Nm_s+m_s^2}\frac{1}{2m_N}\frac{1}{(2\pi)^3}\frac{4\pi}{3}k_f^3=-ig_{\sigma_1N}g_{\sigma_2N}\frac{\rho_{p(n)}}{m_{\sigma_1}}.
\end{align}
From the second to the third
%lines
line, $E_N(\vec{p})$ is approximated by
$m_N$ as before.
In a similar manner,
%the
diagram $(c)$ is evaluated as
\begin{align}
 -i\Sigma_s^{(c)}=&-\tr\int\frac{d^4p}{(2\pi)^4}(-ig_{\sigma_2N})\frac{i(\Slash{p}+m_N)}{p^2-m_N^2}(-ig_{\sigma_1N})(-2\pi)(\Slash{p}+\Slash{q}+m_N)\delta((p+q)^2-m_N^2)\notag\\
 &\times\theta(p_0+q_0)\theta(k_f-|\vec{p}+\vec{q}|)\notag\\
 =&-ig_{\sigma_1N}g_{\sigma_2N}\tr\int\frac{d^4p}{(2\pi)^3}\frac{\Slash{p}-\Slash{q}+m_N}{(p-q)^2-m_N^2}\frac{\Slash{p}+m_N}{2E_N(\vec{p})}\delta(p_0-E_N(\vec{p}))\theta(k_f-|\vec{p}|) \notag\\
 =&-ig_{\sigma_1N}g_{\sigma_2N}\int\frac{d^4p}{(2\pi)^3}\frac{p\cdot(p-q)+m_N^2}{(p-q)^2-m_N^2}\frac{1}{2E_N(\vec{p})}\delta(p_0-E_N(\vec{p}))\theta(k_f-|\vec{p}|) \notag\\
 \sim&-4ig_{\sigma_1N}g_{\sigma_2N}\frac{2m_N^2-m_Nm_s}{-2m_Nm_s+m_s^2}\frac{1}{2m_N}\frac{1}{(2\pi)^3}\frac{4\pi}{3}k_f^{p(n)3}=ig_{\sigma_1N}g_{\sigma_2N}\frac{\rho_{p(n)}}{m_{\sigma_1}} 
\end{align}
From the first to the second
%lines
line, we have changed the integration
variable $p$ to $p-q$.

As in the case of the above calculations, the self-energies
of the pseudoscalar meson from
%the
diagrams $(b)$ and $(c)$ in
Fig.~\ref{diag_self_energy} are evaluated as
\begin{align}
 -i\Sigma_{ps}^{(b)}=&-\tr\int\frac{d^4p}{(2\pi)^4}(g_{\pi_1N}\gamma_5)\frac{i(\Slash{p}+\Slash{q}+m_N)}{(p+q)^2-m_N^2}(g_{\pi_2N}\gamma_5)(\Slash{p}+m_N)(-2\pi)\delta(p^2-m_N^2)\notag\\
 &\times\theta(p_0)\theta(k_f-|\vec{p}|)\notag\\
 =&ig_{\pi_1N}g_{\pi_2N}\int\frac{d^4p}{(2\pi)^3}\frac{4[-p\cdot(p+q)+m_N^2]}{(p+q)^2-m_N^2}\frac{1}{2E_N(\vec{p})}\delta(p_0-E_N(\vec{p}))\theta(k_f-|\vec{p}|) \notag\\
 \sim&ig_{\pi_1N}g_{\pi_2N}\frac{1}{(2\pi)^3}\frac{-4m_Nm_{ps}}{2m_Nm_{ps}+m_{ps}^2}\frac{1}{2m_N}\frac{4\pi}{3}k_f^{p(n)3}\sim-i\frac{g_{\pi_1N}g_{\pi_2N}}{2m_N}\rho_{p(n)}\label{eq_sigma_ps_b} 
\end{align}
and
\begin{align}
  -i\Sigma_{ps}^{(c)}=&-\tr\int\frac{d^4p}{(2\pi)^4}(g_{\pi_2N}\gamma_5)\frac{i(\Slash{p}+m_N)}{p^2-m_N^2}(g_{\pi_1N}\gamma_5)(-2\pi)(\Slash{p}+\Slash{q}+m_N)\delta((p+q)^2-m_N^2)\notag\\
 &\times\theta(p_0+q_0)\theta(k_f-|\vec{p}+\vec{q}|)\notag\\
 =&ig_{\pi_1N}g_{\pi_2N}\int\frac{d^4p}{(2\pi)^3}\frac{4[p\cdot(-p+q)+m_N^2]}{(p-q)^2-m_N^2}\frac{1}{2E_N(\vec{p})}\delta(p_0-E_N(\vec{p}))\theta(k_f-|\vec{p}|) \notag\\
 =&ig_{\pi_1N}g_{\pi_2N}\int\frac{d^4p}{(2\pi)^3}\int\frac{d^4p}{(2\pi)^3}\frac{4p\cdot q}{-2p\cdot q+q^2}\frac{1}{2E_N(\vec{p})}\theta(k_f-|\vec{p}|)\notag\\ 
 \sim&-ig_{\pi_1N}g_{\pi_2N}\frac{1}{(2\pi)^3}\frac{4m_Nm_{ps}}{2m_Nm_{ps}-m_{ps}^2}\frac{1}{2m_N}\frac{4\pi}{3}k_f^{p(n)3}\sim-i\frac{g_{\pi_1N}g_{\pi_2N}}{2m_N}\rho_{p(n)}.\label{eq_sigma_ps_c}
\end{align}
In the last lines of Eqs.~(\ref{eq_sigma_ps_b}) and
(\ref{eq_sigma_ps_c}), we have dropped $m_{ps}$ regarding it as small
compared with the nucleon mass $m_N$.
Thus, the self-energies of mesons, $\Sigma_s(\rho)$ and
$\Sigma_{ps}(\rho)$, are given as 
\begin{align}
 -i\Sigma_s(\rho)=&-i\frac{g_{\sigma_kN}g_{\sigma_k\sigma_1\sigma_2}}{m_{\sigma_k}^2}\rho_{p(n)},\\
 -i\Sigma_{ps}(\rho)=&-i\frac{g_{\sigma_k
 N}g_{\sigma_k\pi_1\pi_2}}{m_{\sigma_k}^2}\rho_{p(n)}-i\frac{g_{\pi_1 N}g_{\pi_2N}}{m_N}\rho_{p(n)}.
\end{align}

Thus, the meson masses in the nuclear medium are given as follows:
%;
\begin{align}
 m_{\pi_3}^2(\rho)=&m_{\pi^\pm}^2(\rho)=m_{\pi^{\pm,0}}^2+2\lambda\sigq\dsigq+2\lambda'(2\sigq\dsigq+\sigs\dsigs)\notag\\
 &+\frac{2}{3}B(2\dsigq+\dsigs)+\frac{g^2\rho}{2m_N}, \\
 m_{\eta_0}^2(\rho)=&m_{\eta_0}^2+\frac{2}{3}\lambda(\sigq\dsigq+2\sigs\dsigs)+2\lambda'(2\sigq\dsigq+\sigs\dsigs)\notag\\
 &+\frac{2}{3}B(2\dsigq+\dsigs)+\frac{g^2\rho}{3m_N}, \\
 m_{\eta_8}^2(\rho)=&m_{\eta_8}^2+\frac{2}{3}\lambda(\sigq\dsigq+2\sigs\dsigs)+2\lambda'(2\sigq\dsigq+\sigs\dsigs)\notag\\
 &-\frac{B}{3}(4\dsigq-\dsigs)+\frac{g^2\rho}{6m_N}, \\
 m_{\eta_0\eta_8}^2(\rho)=&m_{\eta_0\eta_8}^2+\frac{4\lambda}{3\sqrt{2}}(\sigq\dsigq-\sigs\dsigs)-\frac{\sqrt{2}}{3}B(\dsigq-\dsigs)+\frac{g^2\rho}{3\sqrt{2}m_N}, \\ 
 m_{\eta_0\pi_3}^2(\rho)=&m_{\eta_0\pi_3}^2-2\sqrt{\frac{2}{3}}\delta\dsq(2\lambda\sigq-B)-4\sqrt{\frac{2}{3}}\lambda\dsq\dsigq-\frac{g^2\delta\rho}{\sqrt{6}m_N}, \\
 m_{\pi_3\eta_8}^2(\rho)=&m_{\pi_3\eta_8}^2-\frac{4}{\sqrt{3}}\delta\dsq(\lambda\sigq+B)-\frac{4}{\sqrt{3}}\lambda\dsq\dsigq-\frac{g^2\delta\rho}{2\sqrt{3}m_N}.
\end{align}

Due to ISB, the masses of the pseudoscalar mesons are modified as
follows;
\begin{align}
 m_{\pi^0}^2=&m_{\pi_3}^2-\frac{(m_{\pi_3\eta'}^2)^2}{m_{\eta'}^{(is)2}-m_{\pi_3}^2}-\frac{(m_{\pi_3\eta}^2)^2}{m_{\eta}^{(is)2}-m_{\pi_3}^2},\\
 m_{\eta'}^2=&m_{\eta'}^{(is)2}+\frac{(m_{\pi_3\eta'}^2)^2}{m_{\eta'}^{(is)2}-m_{\pi_3}^2},\\
 m_{\eta}^2=&m_{\eta}^{(is)2}+\frac{(m_{\pi_3\eta}^2)^2}{m_{\eta}^{(is)2}-m_{\pi_3}^2}.
\end{align}
The modifications of the meson masses through the mixing of $\pi_3$-$\eta$
or $\pi_3$-$\eta'$ are ignored in this calculation because they are
$O((m_u-m_d)^2)$.
Here, $m_{\eta,\eta'}^{(is)2}$ are the masses of the $\eta,\eta'$ mesons
in the isospin-symmetric limit; $m_{\eta',\eta}^{(is)2}=(m_{\eta_0}^2+m_{\eta_8}^2\pm\sqrt{(m_{\eta_0}^2-m_{\eta_8}^2)^2+4(m_{\eta_0\eta_8}^2)^2})/2$.
$m_{\pi_3\eta}^2$ and $m_{\pi_3\eta'}^2$ are given as
$m_{\pi_3\eta}^2=-\sin\theta_{ps}m_{\pi_3\eta_0}^2+\cos\theta_{ps}m_{\pi_3\eta_8}^2$,
$m_{\pi_3\eta'}^2=\cos\theta_{ps}m_{\pi_3\eta_0}^2+\sin\theta_{ps}m_{\pi_3\eta_8}^2$.

The mixing angles are defined as
\begin{align}
 \begin{pmatrix}
  \pi^0\\
  \eta'\\
  \eta
 \end{pmatrix}=&
\begin{pmatrix}
 \cos\theta_{\pi_3\eta}&&\sin\theta_{\pi_3\eta}\\
 &1 & \\
 -\sin\theta_{\pi_3\eta}& &\cos\theta_{\pi_3\eta} 
\end{pmatrix}
\begin{pmatrix}
 \cos\theta_{\pi_3\eta'}&\sin\theta_{\pi_3\eta}&\\
  -\sin\theta_{\pi_3\eta'}&\cos\theta_{\pi_3\eta'}& \\
 & &1 
\end{pmatrix}
\begin{pmatrix}
1&&\\
 &\cos\theta_{ps}&\sin\theta_{ps}\\
 &-\sin\theta_{ps}&\cos\theta_{ps} 
\end{pmatrix}
\begin{pmatrix}
 \pi_3\\
 \eta_0\\
 \eta_8
\end{pmatrix}\notag\\
 \sim&\begin{pmatrix}
    \pi_3+\sin\theta_{\pi_3\eta'}\eta'^{(is)}+\sin\theta_{\pi_3\eta}\eta^{(is)}\\
    \eta'^{(is)}-\sin\theta_{\pi_3\eta'}\pi_3\\
    \eta^{(is)}-\sin\theta_{\pi_3\eta}\pi_3
 \end{pmatrix}.
\end{align}
From the first to the second line, we omit the $O((m_u-m_d)^2)$ terms.
Here, we set
$\eta'^{(is)}=\cos\theta_{ps}\eta_0+\sin\theta_{ps}\eta_8$ and
$\eta^{(is)}=-\sin\theta_{ps}\eta_0+\cos\theta_{ps}\eta_8$.
The mixing angles between $\eta$-$\pi^0$ and $\eta'$-$\pi^0$ are given
as follows:
%;
\begin{align}
 \tan 2\theta_{ps}=&\frac{2m_{\eta_0\eta_8}^2}{m_{\eta_0}^2-m_{\eta_8}^2},\\
 \tan 2\theta_{\pi^0\eta}=&\frac{2m_{\pi_3\eta}^2}{m_{\pi_3}^2-m_{\eta}^{(is)2}},\\
 \tan 2\theta_{\pi^0\eta'}=&\frac{2m_{\pi_3\eta'}^2}{m_{\pi_3}^2-m_{\eta'}^{(is)2}}.
\end{align}
 
In the same way as before, the mixing of scalar mesons can be written as
follows;
\begin{align}
 \begin{pmatrix}
  a_0^0\\
  \sigma\\
  f_0
 \end{pmatrix}=&
\begin{pmatrix}
 \cos\theta_{\sigma_3f_0}&&\sin\theta_{\sigma_3f_0}\\
 &1 & \\
 -\sin\theta_{\sigma_3f_0}& &\cos\theta_{\sigma_3f_0} 
\end{pmatrix}
\begin{pmatrix}
 \cos\theta_{\sigma_3\sigma}&\sin\theta_{\sigma_3\sigma}&\\
  -\sin\theta_{\sigma_3\sigma}&\cos\theta_{\sigma_3\sigma}& \\
 & &1 
\end{pmatrix}
\begin{pmatrix}
1&&\\
 &\cos\theta_{s}&\sin\theta_{s}\\
 &-\sin\theta_{s}&\cos\theta_{s} 
\end{pmatrix}
\begin{pmatrix}
 \sigma_3\\
 \sigma_0\\
 \sigma_8
\end{pmatrix}\notag\\
 \sim&\begin{pmatrix}
    \sigma_3+\sin\theta_{\sigma_3\sigma}\sigma^{(is)}+\sin\theta_{\sigma_3f_0}f_0^{(is)}\\
    \sigma^{(is)}-\sin\theta_{\sigma_3\sigma}\sigma_3\\
    f_0^{(is)}-\sin\theta_{\sigma_3f_0}\sigma_3
   \end{pmatrix}
\end{align}
The mixing angles are given as follows:
%;
\begin{align}
 m_\sigma^2=&m_\sigma^{(is)2}-\frac{(m_{\sigma_3\sigma}^2)^2}{m_{\sigma_3}^2-m_\sigma^{(is)2}},\\
 m_{f_0}^2=&m_{f_0}^{(is)2}+\frac{(m_{\sigma_3f_0}^2)^2}{m_{f_0}^{(is)2}-m_{\sigma_3}^{(is)2}},\\
 m_{a_0}^2=&m_{\sigma_3}^2+\frac{(m_{\sigma_3\sigma}^2)^2}{m_{\sigma_3}^2-m_\sigma^{(is)2}}-\frac{(m_{\sigma_3f_0}^2)^2}{m_{f_0}^{(is)2}-m_{\sigma_3}^2}, \\
 \tan 2\theta_s=& \frac{2m_{\sigma_0\sigma_8}^2}{m_{\sigma_0}^2-m_{\sigma_8}^2},\\
 \tan 2\theta_{\sigma_3\sigma}=&\frac{2m_{\sigma_3\sigma}^2}{m_{\sigma_3}^2-m_\sigma^{(is)2}}, \\
 \tan 2\theta_{\sigma_3f_0}=&\frac{2m_{\sigma_3f_0}^2}{m_{\sigma_3}^2-m_{\sigma}^{(is)2}}, 
\end{align}
where $m_{\sigma_3\sigma}^2=\cos\theta_s
m_{\sigma_0\sigma_3}^2+\sin\theta_s m_{\sigma_3\sigma_8}^2$ and
$m_{\sigma_3f_0}^2=-\sin\theta_2m_{\sigma_0\sigma_3}^2+\cos\theta_sm_{\sigma_3\sigma_8}^2$.
The corrections of the scalar meson masses from the mixing are ignored in
this calculation, as in the case of the pseudoscalar meson masses.

$m_{ij}^2=\partial^2V_{\rm eff}/\partial \sigma_i\partial \sigma_j\
(i,j=u,d,s)$ appearing in Sect.~\ref{sec_setup} are written as
\begin{align}
 m_{uu}^2=&\mu^2+3\lambda\sigu^2+\lambda'(3\sigu^2+\sigd^2+\sigs^2)\\
 m_{ud}^2=&2\lambda'\sigu\sigd-B\sigs \\
 m_{us}^2=&2\lambda'\sigu\sigs-B\sigd \\
 m_{dd}^2=&\mu^2+3\lambda\sigd^2+\lambda'(\sigu^2+3\sigd^2+\sigs^2) \\
 m_{ds}^2=&2\lambda'\sigd\sigs-B\sigu \\
 m_{ss}^2=&\mu^2+3\lambda\sigs^2+\lambda'(\sigu^2+\sigd^2+3\sigs^2) 
\end{align}

\section{Couplings of mesons}
\label{app_meson_coupling}
In this appendix, we present
%the
explicit forms of the tree-level
couplings of mesons and meson--baryon
appearing in the text and show the calculations of the in-medium
corrections of the vertices.
The modification of the vertices of mesons comes from the diagrams shown
in Fig.~\ref{fig_vertex_1loop}.
Here, the momenta of the outgoing meson $\pi_i$,
incoming mesons $\sigma_f$, and the $\eta$ meson in the initial state
are denoted
%as
by
$k_i$, $\qt=k_1+k_2$, and $q=k_1+k_2+k_3$, respectively.
We write the mass of the meson with the external momenta $k_i$ as $m_i$.
Furthermore,
%we denote
the coupling of the meson $\pi_i$ and nucleon
$g_{\pi_iN}$
%as
is given
%as
by
$g_i$ for short.
The tree-level couplings of mesons are shown in
Appendices~\ref{app_tree_vertex}.
The meson--baryon couplings at
%the
tree level are given in
Appendix~\ref{app_meson_nucl}. 
%In
Appendix~\ref{app_s2ps_vertex} and \ref{app_4ps_vertex}
%,
give the calculation
of the in-medium modification of the scalar--2-pseudoscalar and
the 4-pseudoscalar meson couplings, respectively.

\subsection{Tree-level couplings of mesons}
\label{app_tree_vertex}
Here, we show the tree-level couplings of the mesons.
The 4-pseudoscalar meson and the scalar--2-pseudoscalar meson
couplings at
tree level are obtained as the coefficients of the
$\pi_i\pi_j\pi_k\pi_l$ and $\sigma_f\pi_i\pi_j$ terms in the Lagrangian
of L$\sigma$M given in Eq.~(\ref{eq_lsm_lagrangian}).
The tree-level 4-point couplings of pseudoscalar mesons
$g_{\pi_i\pi_j\pi_k\pi_l}$ are given as
\begin{align}
 g_{\pi^+\pi^-\pi_3\pi_3}=&-(\frac{\lambda}{2}+\lambda'), \\
 g_{\pi^+\pi^-\eta_0\eta_0}=&-(\lambda+\lambda'), \\
 g_{\pi^+\pi^-\eta_0\eta_8}=&-\sqrt{2}\lambda, \\
 g_{\pi^+\pi^-\eta_8\eta_8}=&-\left(\frac{\lambda}{2}+\lambda'\right),
\end{align}
and the scalar--2-pseudosclar couplings $g_{\sigma_i\pi_j\pi_k}$ are
given as follows:
%;
\begin{align}
 g_{\sigma_0\pi^+\pi^-}=&-\left(\frac{\lambda}{\sqrt{3}}(\sigu+\sigd)+\frac{2\lambda'}{\sqrt{3}}(\sigu+\sigd+\sigs)-\frac{B}{\sqrt{3}}\right),\\
 g_{\sigma_3\pi^+\pi^-}=&-\left(\frac{3}{\sqrt{2}}\lambda+\sqrt{2}\lambda'\right)(\sigu-\sigd), \\
 g_{\sigma_8\pi^+\pi^-}=&-\left(\frac{\lambda}{\sqrt{6}}(\sigu+\sigd)+\frac{2}{\sqrt{6}}\lambda'(\sigu+\sigd-2\sigs)+\frac{2}{\sqrt{6}}B\right), \\
 g_{\sigma_0\pi_3\pi_3}=&-\left(\frac{\lambda}{2\sqrt{3}}(\sigu+\sigd)+\frac{\lambda'}{\sqrt{3}}(\sigu+\sigd+\sigs)-\frac{B}{2\sqrt{3}}\right), \\
 g_{\sigma_8\pi_3\pi_3}=&-\left(\frac{\lambda}{2\sqrt{6}}(\sigu+\sigd)+\frac{\lambda'}{\sqrt{6}}(\sigu+\sigd-2\sigs)+\frac{B}{\sqrt{6}}\right), \\
 g_{\sigma_0\eta_0\eta_0}=&-\left(\frac{\lambda}{3\sqrt{3}}(\sigu+\sigd+\sigs)+\frac{\lambda'}{\sqrt{3}}(\sigu+\sigd+\sigs)+\frac{B}{\sqrt{3}}\right), \\
 g_{\sigma_8\eta_0\eta_0}=&-\left(\frac{\lambda}{3\sqrt{6}}(\sigu+\sigd-2\sigs)+\frac{\lambda'}{\sqrt{6}}(\sigu+\sigd-2\sigs)\right), \\
 g_{\sigma_0\eta_8\eta_8}=&-\left(\frac{\lambda}{6\sqrt{3}}(\sigu+\sigd+4\sigs)+\frac{\lambda'}{\sqrt{3}}(\sigu+\sigd+\sigs)-\frac{B}{2\sqrt{3}}\right), \\
 g_{\sigma_8\eta_8\eta_8}=&-\left(\frac{\lambda}{6\sqrt{6}}(\sigu+\sigd-8\sigs)+\frac{\lambda'}{\sqrt{6}}(\sigu+\sigd-2\sigs)-\frac{B}{\sqrt{6}}\right), \\
 g_{\sigma_0\eta_0\eta_8}=&-\frac{2}{3\sqrt{6}}\lambda(\sigu+\sigd-2\sigs), \\
 g_{\sigma_8\eta_0\eta_8}=&-\left(\frac{\lambda}{3\sqrt{3}}(\sigu+\sigd+4\sigs)-\frac{B}{\sqrt{3}}\right), \\
 g_{\sigma_0\pi_3\eta_0}=&-\frac{\sqrt{2}}{3}\lambda(\sigu-\sigd), \\
 g_{\sigma_3\pi_3\eta_0}=&-\left(\frac{\lambda}{\sqrt{3}}(\sigu+\sigd)-\frac{B}{\sqrt{3}}\right), \\
 g_{\sigma_8\pi_3\eta_0}=&-\frac{\lambda}{3}(\sigu-\sigd), \\
 g_{\sigma_0\pi_3\eta_8}=&-\frac{\lambda}{3}(\sigu-\sigd), \\
 g_{\sigma_3\pi_3\eta_8}=&-\left(\frac{\lambda}{\sqrt{6}}(\sigu+\sigd)+\frac{2}{\sqrt{6}}B\right), \\
 g_{\sigma_8\pi_3\eta_8}=&-\frac{\lambda}{3\sqrt{2}}(\sigu-\sigd), \\
 g_{a_0^\pm\eta_0\pi^\mp}=&-\left(\frac{\lambda}{\sqrt{3}}(\sigu+\sigd)-\frac{B}{\sqrt{3}}\right), \\
 g_{a_0^\pm\pi_3\pi^\mp}=&-\left(-\frac{\lambda}{\sqrt{2}}(\sigu-\sigd)\right), \\
 g_{a_0^\pm\eta_8\pi^\mp}=&-\left(\frac{\lambda}{\sqrt{6}}(\sigu+\sigd)+\frac{2}{\sqrt{6}}B\right)
\end{align}

Due to the mixing of the mesons, the couplings are modified as follows:
%;
\begin{align}
 g_{\sigma\pi^0\pi^0}=&\cos\ts g_{\sigma_0\pi_3\pi_3}+\sin\ts g_{\sigma_8\pi_3\pi_3},\\
 g_{f_0\pi^0\pi^0}=&-\sin\ts g_{\sigma_0\pi_3\pi_3}+\cos\ts g_{\sigma_8\pi_3\pi_3},\\
 g_{\sigma\pi^+\pi^-}=&\cos\ts g_{\sigma_0\pi^+\pi^-}+\sin\ts g_{\sigma_8\pi^+\pi^-},\\
 g_{f_0\pi^+\pi^-}=&-\sin\ts g_{\sigma_0\pi^+\pi^-}+\cos\ts g_{\sigma_8\pi^+\pi^-},\\
 g_{\sigma\eta\eta}=&\cos\ts\left(\sin^2\tps g_{\sigma_0\eta_0\eta_0}-\sin
 2\tps g_{\sigma_0\eta_0\eta_8}+\cos^2\tps g_{\sigma_0\eta_8\eta_8}\right)\notag\\
 &+\sin\ts\left(\sin^2\tps g_{\sigma_8\eta_0\eta_0}-\sin
 2\tps g_{\sigma_8\eta_0\eta_8}+\cos^2\tps g_{\sigma_8\eta_8\eta_8}\right),\\
 g_{f_0\eta\eta}=&-\sin\ts\left(\sin^2\tps g_{\sigma_0\eta_0\eta_0}-\sin
 2\tps g_{\sigma_0\eta_0\eta_8}+\cos^2\tps g_{\sigma_0\eta_8\eta_8}\right)\notag\\
 &+\cos\ts\left(\sin^2\tps g_{\sigma_8\eta_0\eta_0}+\sin
 2\tps g_{\sigma_8\eta_0\eta_8}+\cos^2\tps g_{\sigma_8\eta_8\eta_8}\right),\\
 g_{\sigma\eta\eta'}=&\cos\ts\left(-\frac{\sin
 2\tps}{2} g_{\sigma_0\eta_0\eta_0}+\cos
 2\tps g_{\sigma_0\eta_0\eta_8}+\frac{\sin 2\tps}{2}
 g_{\sigma_0\eta_8\eta_8}\right)\notag\\
 &+\sin\ts\left(-\frac{\sin
 2\tps}{2} g_{\sigma_8\eta_0\eta_0}+\cos
 2\tps g_{\sigma_8\eta_0\eta_8}+\frac{\sin 2\tps}{2}
 g_{\sigma_8\eta_8\eta_8}\right),\\
 g_{f_0\eta\eta'}=&-\sin\ts\left(-\frac{\sin
 2\theta_{ps}}{2}g_{\sigma_0\eta_0\eta_0}+\cos
 2\theta_{ps}g_{\sigma_0\eta_0\eta_8}+\frac{\sin
 2\theta_{ps}}{2}g_{\sigma_0\eta_0\eta_8}\right)\notag\\
 &+\cos\ts\left(-\frac{\sin
 2\theta_{ps}}{2}g_{\sigma_8\eta_0\eta_0}+\cos
 2\theta_{ps}g_{\sigma_8\eta_0\eta_8}+\frac{\sin
 2\theta_{ps}}{2}g_{\sigma_8\eta_0\eta_8}\right),\\
 g_{a_0^\pm\pi^\mp\eta}=&-\sin\theta_{ps}g_{a_0^\pm\pi^\mp\eta_0}+\cos\theta_{ps}g_{a_0^\pm\pi^\mp\eta_8}, \\
 g_{a_0^\pm\pi^\mp\eta'}=&\cos\theta_{ps}g_{a_0^\pm\pi^\mp\eta_0}+\sin\theta_{ps}g_{a_0^\pm\pi^\mp\eta_8},\\
 g_{a_0^0\pi^0\eta}=&g_{\sigma_3\pi_3\eta}=-\sin\tps g_{\sigma_3\eta_0\pi_3}+\cos\tps g_{\sigma_3\pi_3\eta_8}
\end{align}
The couplings
%which
that
emerge from ISB are given as follows:
%;
\begin{align}
 g_{a_0\pi^0\pi^0}=&g_{\sigma_3\pi_3\pi_3}+(\sin\tss)g_{\sigma\pi_3\pi_3}+(\sin\tsf)g_{f_0\pi_3\pi_3}\notag\\
 &+2\sin\tpe(-\sin\tps g_{\sigma_3\eta_0\pi_3}+\cos\tps
 g_{\sigma_3\pi_3\eta_8})\notag\\
 &+2\sin\tpep(\cos\tps g_{\sigma_3\eta_0\pi_3}+\sin\tps g_{\sigma_3\pi_3\eta_8}),\\
 g_{a_0^0\pi^+\pi^-}=&g_{\sigma_3\pi^+\pi^-}+\sin\tss
 g_{\sigma\pi^+\pi^-}+\sin\tsf g_{f_0\pi^+\pi^-}, \\
 g_{a_0^\pm\pi^\mp\pi^0}=&g_{a_0^\pm\pi^\mp\pi_3}+\sin\tpe\left(-\sin\tps
 g_{a_0^\pm\pi^\mp\eta_0}+\cos\tps
 g_{a_0^\pm\pi^\mp\eta_8}\right)\notag\\
 &+\sin\tpep\left(\cos\tps
 g_{a_0^\pm\pi^\mp\eta_0}+\sin\tps g_{a_0^\pm\pi^\mp\eta_8}\right), \\
 g_{\sigma\pi^0\eta}=&g_{\sigma\pi_3\eta}
 =\cos\ts(-\sin\tps g_{\sigma_0\pi_3\eta_0}+\cos\tps
 g_{\sigma_0\pi_3\eta_8})\notag\\
 &+\sin\ts(-\sin\tps g_{\sigma_8\pi_3\eta_0}+\cos\tps
 g_{\sigma_8\pi_3\eta_8}), \\
 g_{f_0\pi^0\eta}=&g_{f_0\pi_3\eta}
 =-\sin\ts(-\sin\tps g_{\sigma_0\pi_3\eta_0}+\cos\tps
 g_{\sigma_0\pi_3\eta_8})\notag\\
 &+\cos\ts(-\sin\tps g_{\sigma_8\pi_3\eta_0}+\cos\tps
 g_{\sigma_8\pi_3\eta_8}).
\end{align}

\subsection{Meson--nucleon coupling constants\label{app_meson_nucl}}
The couplings between the meson and nucleon are
%give
given
as follows:
%;
\begin{align}
 g_{\sigma_0N}=&\frac{g}{\sqrt{3}},\\
 g_{\sigma_3N}=&\frac{g}{\sqrt{2}}\tau_3, \\
 g_{a_0^\pm N}=&g, \\
 g_{\sigma_8N}=&\frac{g}{\sqrt{6}},\\
 g_{\eta_0N}=&\frac{g}{\sqrt{3}}\gamma_5,\\
 g_{\pi_3N}=&\frac{g}{\sqrt{2}}\gamma_5\tau_3, \\
 g_{\pi^\pm N}=&g\gamma_5, \\
 g_{\eta_8N}=&\frac{g}{\sqrt{6}}\gamma_5, \\
 g_{\sigma N}=&\cos\ts g_{\sigma_0N}+\sin\tps g_{\sigma_8N}, \\
 g_{f_0N}=&-\sin\ts g_{\sigma_0N}+\cos\ts g_{\sigma_8N}, \\
 g_{\eta' N}=&\cos\tps g_{\eta_0N}+\sin\tps g_{\eta_8N}, \\
 g_{\eta N}=&-\sin\tps g_{\eta_0N}+\cos\tps g_{\eta_8N}. 
\end{align}

\subsection{Scalar--2-pseudoscalar meson couplings}
\label{app_s2ps_vertex}
First, we evaluate $\Gamma_{(\alpha)}$ from
%the
diagram ($\alpha$) in
Fig.~\ref{fig_vertex_1loop}.
We divide $\Gamma_{(\alpha)}$ into three parts:
%;
$\Gamma_{(\alpha)}^{(1)}$, $\Gamma_{(\alpha)}^{(2)}$, and
$\Gamma_{(3)}^{(3)}$ come from the diagrams with the nuclear hole state
between $\sigma_f$ and $\pi_1$, $\pi_1$ and $\pi_2$, and $\pi_2$ and
$\sigma_f$, respectively. 
$\Gamma_{(\alpha)}^{(1)}$ is evaluated as follows:
\begin{align}
 i\Gamma_\alpha^{(1)}(\rho)=&-\tr\int\frac{d^4p}{(2\pi)^4}(-ig_{\sigma_f
 N})(-2\pi)(\Slash{\qt}+\Slash{p}+m_N)\delta((\qt+p)^2-m_N^2)\theta(p_0+\qt_0)\theta(k_f-|\vec{\qt}+\vec{p}|)\notag\\
 &\times(g_1\gamma_5)\frac{i(\Slash{p}+\Slash{k}_2+m_N)}{(p+k_1)^2-m_N^2}(g_2\gamma_5)\frac{i(\Slash{p}+m_N)}{p^2-m_N^2}\notag\\
 =&ig_{\sigma_f N}g_1g_2\tr\int\frac{d^4p}{(2\pi)^3}\frac{(\Slash{p}+m_N)(-\Slash{p}+\Slash{k}_1+m_N)(\Slash{p}-\Slash{\qt}+m_N)}{((p-k_1)^2-m_N^2)((p-\qt)^2-m_N^2)}\theta(p_0)\delta(p^2-m_N^2)\theta(k_f-\left|\vec{p}\right|)\notag\\
 =&ig_{\sigma_f N}g_1g_2\tr\int\frac{d^4p}{(2\pi)^3}\frac{4m_N(2p_1\cdot
 k_1-k_1^2-k_1\cdot k_2)}{((p-k_1)^2-m_N^2)((p-\qt)^2-m_N^2)}\frac{1}{2E_N(\vec{p})}\delta(p_0-E_N(\vec{p}))\theta(k_f-\left|\vec{p}\right|) \notag\\
 \sim&ig_{\sigma_f N}g_1g_2\frac{2m_NE_1-m_1^2-k_1\cdot
 k_2}{(2m_NE_1-m_1^2)(2m_N(E_1+E_2)-(k_1+k_2)^2)}\rho_{p(n)}\notag\\
 \sim&i\frac{g_{\sigma_fN}g_1g_2}{2m_N(E_1+E_2)}\rho_{p(n)}.
\end{align}
The meson--nucleon couplings $g_{\sigma_fN}$ and $g_i$ are given in
Appendix \ref{app_meson_nucl}.
Using the same approximations as the calculation of scalar meson
self-energy performed in Appendix~\ref{app_in_med_mass},
we take only the leading-order contribution with respect to
$k_f$ and the external momenta $k_i$ regarding them as small compared
with the other quantities.
From the third to the fourth
%lines
line, we use the approximation where
$E_N(\vec{p})$ is taken as $m_N$.
In much the same way as the above calculation, $\Gamma_\alpha^{(2)}(k_f)$ and
$\Gamma_\alpha^{(3)}(k_f)$
are written as 
\begin{align}
 i\Gamma_\alpha^{(2)}(\rho)=&-\tr\int\frac{d^4p}{(2\pi)^4}(-ig_{\sigma_f
 N})\frac{i}{\Slash{\qt}+\Slash{p}-m_N}(g_1\gamma_5)(-2\pi)(\Slash{p}+\Slash{k}_2+m_N)\notag\\
 &\times\theta(p_0+k_{20})\delta((p+k_2)^2-m_N^2)\theta(k_f-|\vec{p}+\vec{k}_2|)(g_2\gamma_5)\frac{i}{\Slash{p}-m_N}\notag\\
 =&ig_{\sigma_f N}g_1g_2\tr\int\frac{d^4p}{(2\pi)^3}\frac{(\Slash{p}+\Slash{k}_1+m_N)(-\Slash{p}+m_N)(\Slash{p}-\Slash{k}_2+m_N)}{((p+k_1)^2-m_N^2)((p-k_2)^2-m_N^2)}\notag\\
 =&ig_{\sigma_f N}g_1g_2\int\frac{d^4p}{(2\pi)^3}\frac{-4m_Nk_1\cdot
 k_2}{(2p\cdot k_1+m_1^2)(-2p\cdot k_2+m_2^2)}\notag\\
 \sim&ig_{\sigma_f N}g_1g_2\frac{k_1\cdot
 k_2}{(2m_NE_1+m_1^2)(2m_NE_2-m_2^2)}\rho_{p(n)}\sim
 i\frac{g_{\sigma_fN}g_1g_2}{4m_N^2E_1E_2}(k_1\cdot k_2)\rho_{p(n)}. 
\end{align}
and
\begin{align}
 i\Gamma_\alpha^{(3)}(\rho)=&-\tr\int\frac{d^4p}{(2\pi)^4}(-ig_{\sigma_f
 N})\frac{i}{\Slash{\qt}+\Slash{p}-m_N}g_1\gamma_5\frac{i}{\Slash{p}+\Slash{k}_2-m_N}g_2\gamma_5(-2\pi)(\Slash{p}+m_N)\notag\\
 &\times\delta(p^2-m_N^2)\theta(p_0)\theta(k_f-|\vec{p}|)\notag\\
 =&ig_{\sigma_f
 N}g_1g_2\int\frac{d^4p}{(2\pi)^3}\frac{\tr(\Slash{p}+\Slash{\qt}+m_N)(-\Slash{p}-\Slash{k}_2+m_N)(\Slash{p}+m_N)}{\left((p+\qt)^2-m_N^2\right)\left((p+k_2)^2-m_N^2\right)}\frac{1}{2E_N(\vec{p})}\theta(p_0)\theta(k_f-|\vec{p}|)\notag\\
 =&ig_{\sigma_f
 N}g_1g_2\int\frac{d^4p}{(2\pi)^3}\frac{-4m_N(2p+\qt)\cdot
 k_2}{\left((p+\qt)^2-m_N^2\right)\left((p+k_2)^2-m_N^2\right)}\frac{1}{2E_N(\vec{p})}\theta(p_0)\theta(k_f-|\vec{p}|)\notag\\
 =&ig_{\sigma_f
 N}g_1g_2\int\frac{d^4p}{(2\pi)^3}\frac{-4m_N(2p+k_1+k_2)\cdot
 k_2}{\left(2p\cdot(k_1+k_2)+(k_1+k_2)^2\right)\left(2p\cdot k_2+k_2^2\right)}\frac{1}{2E_N(\vec{p})}\theta(p_0)\theta(k_f-|\vec{p}|)\notag\\
 \sim&-ig_{\sigma_f N}g_1g_2\frac{2m_NE_2+k_1\cdot
 k_2+m_{\pi_2}^2}{(2m_N(E_1+E_2)+(k_1+k_2)^2)(2m_NE_2+m_{\pi_2}^2)}\rho_{p(n)}\notag\\
 &\sim-i\frac{g_{\sigma_fN}g_{\pi_1N}g_{\pi_2N}}{2m_N(E_1+E_2)}\rho_{p(n)}. 
\end{align}
Thus, the correction of the couplings of mesons in the
symmetric nuclear medium $i\Gamma_\alpha=i\Gamma_\alpha^{(1)}+i\Gamma_\alpha^{(2)}+i\Gamma_\alpha^{(3)}$ is given by
\begin{align}
 i\Gamma_\alpha(\rho)=&i\frac{g_{\sigma_fN}g_1g_2}{4m_N^2E_1E_2}(k_1\cdot
 k_2)\rho_{p(n)}. \label{eq_gamma_3pt}
\end{align}

\subsection{4-pseudoscalar meson vertex in the nuclear medium}
\label{app_4ps_vertex}
Next, we evaluate
%the
diagram $(\beta)$ in Fig.~\ref{fig_vertex_1loop}.
We write
$i\Gamma_\beta=i\Gamma_\beta^{(1)}+i\Gamma_\beta^{(2)}+i\Gamma_\beta^{(3)}+i\Gamma_\beta^{(4)}$.
For $\Gamma_\beta^{(1)}$, the propagator between
%$\sigma_f$
$\eta$
and $\pi_1$
is the nuclear hole state.
$\Gamma_\beta^{(1)}(q)$ is written as
\begin{align}
 i\Gamma_\beta^{(1)}(q)=&-\tr\int\frac{d^4p}{(2\pi)^4}(g_{\eta N}\gamma_5)(-2\pi)(\Slash{p}+\Slash{q}+m_N)\delta((p+q)^2-m_N^2)\theta(k_f-|\vec{p}+\vec{q}|)(g_1\gamma_5)\notag\\
 &\times\frac{i(\Slash{p}+\Slash{k}_2+\Slash{k}_3+m_N)}{(p+k_2+k_3)^2-m_N^2}(g_2\gamma_5)\frac{i(\Slash{p}+\Slash{k}_3+m_N)}{(p+k_3)^2-m_N^2}(g_3\gamma_5)\frac{i(\Slash{p}+m_N)}{p^2-m_N^2} \notag\\
 =&-ig_{\eta N}g_1g_2g_3\tr\int\frac{d^3p}{(2\pi)^3}\frac{(-\Slash{p}-\Slash{q}+m_N)(\Slash{p}+\Slash{k}_2+\Slash{k}_3+m_N)(-\Slash{p}-\Slash{k}_3+m_N)(\Slash{p}+m_N)}{((p+k_2+k_3)^2-m_N^2)((p+k_3)^2-m_N^2)(p^2-m_N^2)}\notag\\
 &\times\delta((p+q)^2-m_N^2)\theta(p_0+q_0)\theta(k_f-\left|\vec{p}+\vec{q}\right|)\notag\\
 =&-ig_{\eta N}g_1g_2g_3\tr\int\frac{d^4p}{(2\pi)^3}\frac{(-\Slash{p}+m_N)(\Slash{p}-\Slash{k}_1+m_N)(-\Slash{p}+\Slash{k}_1+\Slash{k}_2+m_N)(\Slash{p}-\Slash{q}+m_N)}{[(p-k_1)^2-m_N^2][(p-k_1-k_2)^2-m_N^2][(p-q)^2-m_N^2]}\theta(p_0)\notag\\
 &\times\delta(p^2-m_N^2)\theta(k_f-|\vec{p}|)\notag\\ 
 \sim&ig_1g_2g_3g_4\frac{2m_NE_1-k_1\cdot(k_1+k_2)}{(2m_NE_1-m_1^2)(2m_N(E_1+E_2)-(k_1+k_2)^2)(2m_N-m_\eta)}\rho_{p(n)}\notag\\
 \sim&i\frac{g_{\eta N}g_1g_2g_3}{4m_N^2(E_1+E_2)}\rho_{p(n)}.
\end{align}
Here, the approximation taking $E_N(\vec{p})$ as $m_N$ is used as in the
case of the above calculation.
$\Gamma_\beta^{(2)}(q)$, which is the contribution from the diagram with
the nuclear hole state between $\pi_1$ and $\pi_2$, is evaluated as
\begin{align}
 i\Gamma_\beta^{(2)}(q)=&-\tr\int\frac{d^4p}{(2\pi)^4}(g_{\eta N}\gamma_5)\frac{i(\Slash{p}+\Slash{q}+m_N)}{(p+q)^2-m_N^2}(g_1\gamma_5)(-2\pi)(\Slash{p}+\Slash{k}_2+\Slash{k}_3+m_N)\delta((p+k_2+k_3)^2-m_N^2)\notag\\
&\times\theta(p_0+k_{20}+k_{30})\theta(k_f-|\vec{p}+\vec{k}_2+\vec{k_3}|)(g_2\gamma_5)\frac{i(\Slash{p}+\Slash{k}_3+m_N)}{(p+k_3)^2-m_N^2}(g_3\gamma_5)\frac{i(\Slash{p}+m_N)}{p^2-m_N^2}\notag\\
 =&-ig_{\eta N}g_1g_2g_3\tr\int\frac{d^4p}{(2\pi)^3}\frac{-\Slash{p}-\Slash{k}_1+m_N}{(p+k_1)^2-m_N^2}(\Slash{p}+m_N)\delta(p^2-m_N^2)\theta(p_0)\theta(k_f-|\vec{p}|)\notag\\
 &\times\frac{-\Slash{p}+\Slash{k}_2+m_N}{(p-k_2)^2-m_N^2}\frac{\Slash{p}-\Slash{k}_2-\Slash{k}_3+m_N}{(p-k_2-k_3)^2-m_N^2}\notag\\ 
 \sim&ig_{\eta
 N}g_1g_2g_3\frac{2m_NE_1E_2-E_1k_2\cdot(k_2+k_3)-E_2(k_1\cdot
 k_3)+E_3(k_1\cdot
 k_2)}{(2m_NE_1+m_1^2)(2m_NE_2-m_2^2)(2m_N(E_2+E_3)-(k_2+k_3)^2)}\rho_{p(n)}\notag\\
 \sim&i\frac{g_{\eta N}g_1g_2g_3}{4m_N^2(E_2+E_3)}\rho_{p(n)}. 
\end{align}
$\Gamma_\beta^{(3)}(q)$ with the hole state between $\pi_2$
and $\pi_3$ is given by
\begin{align}
 i\Gamma_\beta^{(3)}(q)=&-\tr\int\frac{d^4p}{(2\pi)^4}(g_{\eta N}\gamma_5)\frac{i(\Slash{p}+\Slash{q}+m_N)}{(p+q)^2-m_N^2}(g_1\gamma_5)\frac{i(\Slash{p}+\Slash{k}_2+\Slash{k}_3+m_N)}{(p+k_2+k_3)^2-m_N^2}(g_2\gamma_5)(-2\pi)\notag\\
 &\times(\Slash{p}+\Slash{k}_3+m_N)\delta((p+k_3)^2-m_N^2)\theta(p_0+k_{30})\theta(k_f-|\vec{p}+\vec{k}_3|)(g_3\gamma_5)\frac{i(\Slash{p}+m_N)}{p^2-m_N^2}\notag\\
 =&-ig_{\eta N}g_1g_2g_3\tr\int\frac{d^4p}{(2\pi)^3}\frac{-\Slash{p}-\Slash{k}_1-\Slash{k}_2+m_N}{(p+k_1+k_2)^2-m_N^2}\frac{\Slash{p}+\Slash{k}_2+m_N}{(p+k_2)^2-m_N^2}(-\Slash{p}+m_N)\notag\\
 &\times\delta(p^2-m_N^2)\theta(p_0)\theta(k_f-|\vec{p}|)\frac{\Slash{p}-\Slash{k}_3+m_N}{(p-k_3)^2-m_N^2} \notag\\
 \sim&-ig_{\eta N}g_1g_2g_3\frac{2m_NE_2E_3-E_1(k_2\cdot k_3)+E_2(k_1\cdot
 k_3)+E_3k_2\cdot(k_1+k_2)}{[2m_N(E_1+E_2)+(k_1+k_2)^2][2m_NE_2+m_2^2][2m_NE_3-m_3^2]}\notag\\
 \sim&-i\frac{g_{\eta N}g_1g_2g_3}{4m_N^2(E_1+E_2)}\rho_{p(n)}. 
\end{align}
Finally, $i\Gamma_\beta^{(4)}(q)$,
which is the contribution from the diagram with the hole state between
$\pi_3$ and
%$\sigma_f$
$\eta$,
is written as
\begin{align}
 i\Gamma_\beta^{(4)}(q)=&-\tr\int\frac{d^4p}{(2\pi)^4}(g_{\eta N}\gamma_5)\frac{i(\Slash{p}+\Slash{q}+m_N)}{(p+q)^2-m_N^2}(g_1\gamma_5)\frac{i(\Slash{p}+\Slash{k}_2+\Slash{k}_3+m_N)}{(p+k_2+k_3)^2-m_N^2}(g_2\gamma_5)\frac{i(\Slash{p}+\Slash{k}_3+m_N)}{(p+k_3)^2-m_N^2}\notag\\
&\times(g_3\gamma_5)(-2\pi)(\Slash{p}+m_N)\delta(p^2-m_N^2)\theta(p_0)\theta(k_f-|\vec{p}|)\notag\\
 =&-ig_{\eta N}g_1g_2g_3\tr\int\frac{d^4p}{(2\pi)^3}\frac{-\Slash{p}-\Slash{q}+m_N}{(p+q)^2-m_N^2}\frac{\Slash{p}+\Slash{k}_2+\Slash{k}_3+m_N}{(p+k_2+k_3)^2-m_N^2}\frac{-\Slash{p}-\Slash{k}_3+m_N}{(p+k_3)^2-m_N^2}\notag\\
 &\times(\Slash{p}+m_N)\delta(p^2-m_N^2)\theta(p_0)\theta(k_f-|\vec{p}|)\notag\\
 \sim&-ig_{\eta N}g_1g_2g_3\frac{2m_NE_3+k_3\cdot(k_2+
 k_3)}{(2m_N+m_\eta)(2m_N(E_2+E_3)+(k_2+k_3)^2)(2m_NE_3+m_3^2)}\rho_{p(n)}\notag  \\
 \sim&-i\frac{g_{\eta N}g_1g_2g_3}{4m_N^2(E_2+E_3)}\rho_{p(n)}. 
\end{align}

From the above calculations,
$i\Gamma_\beta(q)=i\Gamma_\beta^{(1)}+i\Gamma_\beta^{(2)}+i\Gamma_\beta^{(3)}+i\Gamma_\beta^{(4)}$
vanishes in this calculation.

\section{Nucleon self-energy in the nuclear medium}
\label{app_nucl_mass}
 We evaluate the contribution from the diagram in Fig.~\ref{diag_nucl_fock}
to the self-energy $\Sigma_N$ of the nucleon in the nuclear medium,
 which reads
%\sout{The contribution from the diagram shown in
%Fig.~\ref{diag_nucl_fock} $\Sigma_N$ is written as follows;}
\begin{align}
 -i\Sigma_N=&\int\frac{d^4p}{(2\pi)^4}(-ig_{\sigma_fN})(-2\pi)(\Slash{p}+\Slash{q}+m_N)\theta(p_0+q_0)\delta((p+q)^2-m_N^2)\notag\\
 &\times\theta(k_f-|\vec{p}+\vec{q}|)(-ig_{\sigma_fN})\frac{i}{p^2-m_{\sigma_f}^2}\notag\\
 =&ig_{\sigma_fN}^2\int\frac{d^4p}{(2\pi)^3}(\Slash{p}+m_N)\theta(p_0)\delta(p^2-m_N^2)\theta(k_f-|\vec{p}|)\frac{1}{(p-q)^2-m_N^2}\notag\\
 =&\left.ig_{\sigma_fN}^2\int\frac{d^3p}{(2\pi)^3}(\Slash{p}+m_N)\frac{1}{2E_N(\vec{p})}\frac{1}{2m_N^2-2m_NE_N(\vec{p})-m_{\sigma_f}^2}\right|_{p_0=E_N(\vec{p})}.\notag
\end{align}
Setting the incoming nucleon momentum
% \sout{$q$} 
as $q=(m_N,\vec{0})$ and
ignoring the higher-order terms with respect to $k_f$, 
$\Sigma_N$ is
obtained as 
\begin{align}
 -i\Sigma_N=&ig_{\sigma_fN}^2\int\frac{d^3p}{(2\pi)^3}(2m_N)\frac{1}{2m_N}\frac{1}{-m_{\sigma_f}^2}\\
 =&-i\frac{g_{\sigma_fN}^2}{2m_{\sigma_f}^2}\rho_{p(n)}. 
\end{align}

\bibliographystyle{aip}
\bibliography{ref}

\end{document}